\documentclass[aoas]{imsart}

\RequirePackage{amsthm,amsmath,amsfonts,amssymb}
\RequirePackage[authoryear]{natbib}
\RequirePackage[colorlinks,citecolor=blue,urlcolor=blue]{hyperref}
\RequirePackage{graphicx}
\graphicspath{{Results/}}

\usepackage{subfigure}
\usepackage{rotating}
\usepackage{booktabs}
\usepackage{multirow}
\usepackage{bm}

\newcommand\hl[1]{#1}

\begin{document}

\begin{frontmatter}
\title{Sophisticated and small versus simple and sizeable: When does it pay off to introduce drifting coefficients in Bayesian vector autoregressions?}
\runtitle{Sophisticated and small versus simple and sizeable}

\vspace{2em}
This is a preprint of the version published in:\\\emph{Journal of Forecasting}\\\url{https://doi.org/10.1002/for.3121}
\vspace{1em}


\begin{aug}
\author[A]{\fnms{Martin}~\snm{Feldkircher}\orcid{0000-0002-5511-9215}\ead[label=e1]{martin.feldkircher@da-vienna.ac.at}},
\author[B]{\fnms{Luis}~\snm{Gruber}\orcid{0000-0002-2399-738X}\ead[label=e2]{luis.gruber@aau.at}},
\author[C]{\fnms{Florian}~\snm{Huber}\orcid{0000-0002-2896-7921}\ead[label=e3]{florian.huber@plus.ac.at}}
\and
\author[B]{\fnms{Gregor}~\snm{Kastner}\thanks{Corresponding author.}\corref{} \orcid{0000-0002-8237-8271}\ead[label=e4]{gregor.kastner@aau.at}}
\address[A]{Vienna School of International Studies (DA)\printead[presep={,\ }]{e1}}
\address[B]{Department of Statistics, University of Klagenfurt, Universit\"{a}tsstra\ss e 65-67, 9020 Klagenfurt am W\"{o}rthersee, Austria\printead[presep={,\ }]{e2,e4}}
\address[C]{University of Salzburg\printead[presep={,\ }]{e3}}
\end{aug}

\begin{abstract}
We assess the relationship between model size and  complexity in the time-varying parameter vector autoregression framework via thorough predictive exercises for the euro area, the United Kingdom, and the United States. It turns out that sophisticated dynamics through drifting coefficients are important in small data sets, while simpler models tend to perform better in sizeable data sets. To combine the best of both worlds, novel shrinkage priors help to mitigate the curse of dimensionality, resulting in competitive forecasts for all scenarios considered. Furthermore, we discuss dynamic model selection to improve upon the best performing individual model for each point in time.
\end{abstract}

\begin{keyword}
\kwd{global-local shrinkage priors}
\kwd{density predictions}
\kwd{hierarchical modeling}
\kwd{stochastic volatility}
\kwd{dynamic model selection}
\end{keyword}
\end{frontmatter}

\section{Introduction}
In predictive inference, two main bearings can be found. First, simple models are increasingly replaced by more sophisticated versions in order to avoid functional misspecification. Second, due to increased data availability,  information sets become more sizeable, and models thus higher dimensional, which in turn decreases the likelihood of omitted variable bias. The goal of this paper is a systematic assessment of the relationship between model size and complexity using different macroeconomic data sets by comparing the predictive performance of multivariate time series models that range from being relatively simple (i.e., feature constant parameters and heteroskedastic shocks) to very flexible (time-varying parameters and heteroskedastic shocks). Our conjecture is that the introduction of drifting coefficients  in multivariate time series regressions can control for an omitted variable bias in small-scale models or, conversely, larger information sets can substitute for non-linear model dynamics. 

\hl{The choice between model size and model complexity is not innocuous. Large models, such as vector autoregressions (VARs) with many endogenous variables, naturally avoid an omitted variable bias. This often translates into superior predictive performance  \citep{banbura2010} and avoids  puzzles commonly observed in empirical macroeconomics \citep[such as the price puzzle; see][]{sims1992interpreting}. Adding stochastic volatility (SV) often further improves forecast performance \citep{clark2011real, Clark2015} but only slightly increases the computational burden if state-of-the-art sampling algorithms are used \citep[for recent examples, see][]{kastner2020sparse, chan2023largeorder}.}

\hl{Both VARs and VARs with SV, however, are  over-parameterized if the number of endogenous variables becomes moderately large and can be miss-specified if the data generating process (DGP) features nonlinearities. This calls for appropriate regularization techniques, and the Bayesian literature has proposed appropriate shrinkage priors to overcome overfitting issues. Particular examples are the popular class of Minnesota-type priors \citep{Doan1984, Litterman1986, Sims1998}, stochastic search variable selection  priors \citep{george1993variable, george2008bayesian}, and global-local shrinkage priors \citep[see, e.g.,][]{polson2010shrink, Brown2010griffin,bhattacharya2015dirichlet, Zhang2022, Huber2017,FOLLETT2019,  kastner2020sparse}. More recently, approaches combining the merits of Minnesota-type shrinkage priors and global-local priors have been developed and used successfully in constant-parameter VARs \citep[see, e.g.,][]{cross2020macroeconomic, chan2021minnesota, gruber2022sparsedense}.}

Complicated models, by contrast, are often difficult to estimate and interpret and do not scale well to large data sets. In particular,  time-varying parameter  VARs (TVP-VARs) are flexible models that, as shown in \cite{granger2008non}, allow for capturing nonlinearities of unknown form. In macroeconomic data, these nonlinearities often relate to parameter change and structural breaks over longer time periods. However, TVP-VARs are often applied to small data sets with only a handful of endogenous variables.  The main reason is that the computational burden increases appreciable with increasing model dimension. However, there exists a large body of literature \citep[see, e.g.,][]{d2013macroeconomic, huber2021inducing} that shows that small to moderately sized TVP-VARs can compete with large-dimensional VARs in predictive performance. In recent papers, researchers increasingly adopt shrinkage priors in TVP regressions to force the model toward a time-invariant specification \citep[see, e.g.,][]{belmonte2014hierarchical,KALLI2014, cadonna2020, huber2021inducing}. These papers leverage the non-centered parameterization of the state space model \citep[see][]{fruhwirth2010stochastic} to disentangle the time-invariant component of the model from the dynamic part. \hl{In another recent paper, \cite{chan2023largehybrid} provides methods that allow for estimating TVP-VARs where some equations feature TVPs, while other equations are characterized by constant coefficients.}

To sum up this discussion, applied researchers face a trade-off. On the one hand, using large data sets alleviates the risk of missing important information that might be relevant for macroeconomic forecasting. On the other hand, estimating sophisticated models such as TVP-VARs in large dimensions is extremely challenging, and the risk of overfitting is further intensified. Table \ref{tab:overview} provides a summary of the pros and cons of these different general modeling strategies. 

\begin{table}[t!]
 \caption{Trade-offs in VAR modeling}
 \label{tab:overview}
 \centering
 \begin{tabular}{rr|cc}
&  &  \multicolumn{2}{c}{\hspace{2cm}\textbf{Model}} \\
&  &  simple & sophisticated \\
\hline
   \multirow{2}{*}{\textbf{Information set}} & Small & Omitted variables + misspecification &  Omitted variables\\
   & Sizeable & Misspecification & Overfitting
 \end{tabular}
\end{table}

In this paper, we explore this relationship empirically by considering three well-known data sets for the  euro area (EA), the United Kingdom (UK) and the United States (US). We consider VARs that differ in terms of their model dimensions. Specifically, we consider  for each data set a small model that features 3 variables, a moderately sized one with 7 variables and a large model with 15 variables. To focus on the relationship between model size and complexity, we differentiate between models that feature constant and drifting parameters. In both specifications, we assume that the shocks are heteroskedastic and follow an SV model. To control for overfitting, two recent shrinkage priors, the Normal-Gamma (NG) prior \citep{griffin2010inference} and the Dirichlet-Laplace (DL) prior \citep{bhattacharya2015dirichlet}, are used to induce shrinkage in our different model specifications.

Our results are threefold: First, we show that the proposed TVP-VAR-SV shrinkage models improve one-step ahead forecasts.  Allowing for time variation and using shrinkage priors leads to smaller drops in forecast performance during the global financial crisis -- a finding that is also corroborated by looking at model weights in a dynamic model selection exercise. Second,  comparing the proposed priors, we find that the DL prior shows a strong performance in small-scale applications, while the NG prior outperforms using larger information sets. This is driven by the higher degree of shrinkage the NG prior provides, which is especially important for large scale applications. Last, we demonstrate that the larger the information set, the stronger the forecast performance of a simple, constant parameter VAR with SV.  However, also here, the NG-VAR-SV model turns out to be a valuable alternative, providing forecasts that are not far off those of the constant parameter competitor. To allow for different models at different points in time, we also discuss the possibility of dynamic model selection \citep{Koop2009}.

The remainder of the paper is structured as follows. The second section sets the stage, introduces a standard TVP-VAR-SV model, and highlights typical estimation issues involved. Section 3 describes in detail the prior setup adopted. Section 4 presents the necessary details to estimate the model, including an overview of the Markov chain Monte Carlo (MCMC) algorithm and the relevant conditional posterior distributions. Section 5 provides empirical results alongside the main findings of our forecasting comparison. Furthermore, it contains a discussion of dynamic model selection. Finally, the last section summarizes and concludes the paper.

\section{Econometric framework}
In this paper, the model of interest is a TVP-VAR with (SV) in the spirit of \cite{primiceri2005time}. The model summarizes the joint dynamics of an $M$-dimensional zero-mean vector of macroeconomic time series $\{\boldsymbol{y}_t\}_{t=1}^T$ as follows:\footnote{To simplify the model exposition, we omit an intercept term in this section. Irrespectively of this, we allow for non-zero intercepts in the empirical applications that follow.}
\begin{equation}
\boldsymbol{y}_t = \boldsymbol{A}_{1t} \boldsymbol{y}_{t-1}+\dots+\boldsymbol{A}_{pt} \boldsymbol{y}_{t-p} + \boldsymbol{\varepsilon}_t,~\boldsymbol{\varepsilon}_t \sim \mathcal{N}(\boldsymbol{0}_M, \boldsymbol{\Sigma}_t). \label{eq: obs1}
\end{equation}
The $M \times M$ matrix $\boldsymbol{A}_{jt}$ ($j=1,\dots,p$) contains time-varying autoregressive coefficients, $\boldsymbol{\varepsilon}_t$ is a vector white noise error with zero mean and a time-varying variance-covariance matrix $\boldsymbol{\Sigma}_t= \boldsymbol{H}_t \boldsymbol{V}_t \boldsymbol{H}_t'$. $\boldsymbol{H}_t$ is a lower unitriangular matrix,  and $\boldsymbol{V}_t=\text{diag}(e^{v_{1t}},\dots, e^{v_{Mt}})$ denotes a diagonal matrix with time-varying shock variances. The model in \autoref{eq: obs1} can be cast in a standard regression form as follows,
\begin{equation}
\boldsymbol{y}_t = \boldsymbol{A}_t \boldsymbol{x}_t + \boldsymbol{\varepsilon}_t, \label{eq: obs2}
\end{equation}
with $\boldsymbol{A}_t = (\boldsymbol{A}_{1t}, \dots, \boldsymbol{A}_{pt})$ being an $M\times (pM)$ matrix and $\boldsymbol{x}_t = (\boldsymbol{y}'_{t-1},\dots, \boldsymbol{y}'_{t-p})'$. Following \cite{cogley2005drifts}, we can rewrite \autoref{eq: obs2} as
\begin{equation}
\boldsymbol{y}_t - \boldsymbol{A}_t \boldsymbol{x}_t = \boldsymbol{H}_t \boldsymbol{\eta}_t, \text{ with } \boldsymbol{\eta}_t \sim \mathcal{N}(\boldsymbol{0}, \boldsymbol{V}_t),
\end{equation}
and multiplying from the left with $\tilde{\boldsymbol{H}}_t := \boldsymbol{H}_t^{-1}$ yields
\begin{equation}
\tilde{\boldsymbol{H}}_t  \boldsymbol{\varepsilon}_t = \boldsymbol{\eta}_t.
\end{equation}
For further illustration, note that the first two equations of the system are given by 
\begin{align}
\varepsilon_{1t} &= \eta_{1t},\\
\tilde{h}_{2 1,t}  \varepsilon_{1t} + \varepsilon_{2t}&= \eta_{2t}, \label{secondeq}
\end{align}
with $\tilde{h}_{2 1,t}$ denoting the second element of the first column of $\tilde{\boldsymbol{H}}_t$. \autoref{secondeq} can be rewritten as
\begin{equation}
 y_{2t} = \boldsymbol{A}_{2 \bullet, t} \boldsymbol{x}_t - \tilde{h}_{2 1,t}  \varepsilon_{1t} + \eta_{2t}, \label{secondeqaug}
\end{equation}
where $\boldsymbol{A}_{i \bullet, t}$ denotes the $i$th row of $\boldsymbol{A}_t$. More generally, the $i$th equation of the system is a standard regression model augmented with the residuals of the preceding $i-1$ equations,
\begin{equation}
y_{it} = \boldsymbol{A}_{i \bullet, t} \boldsymbol{x}_t - \sum_{s=1}^{i-1} \tilde{h}_{i s, t} \varepsilon_{st} + \eta_{it}.
\end{equation}
Thus, the $i$th equation is a standard regression model with $K_i = pM + i-1$ explanatory variables given by $\boldsymbol{z}_{it} = ( \boldsymbol{x}_t', -\varepsilon_{1t}, \dots, -\varepsilon_{i-1,t})'$ and a $K_i$-dimensional  time-varying coefficient vector $\boldsymbol{B}_{it}= ( \boldsymbol{A}_{i\bullet,t}, \tilde{h}_{i1, t}, \dots, \tilde{h}_{i i-1, t})'$.  
For each equation $i>1$, the corresponding dynamic regression model is then given by
\begin{equation}
y_{it} = \boldsymbol{B}'_{it} \boldsymbol{z}_{it} + \eta_{it}. \label{eq: regression_i}
\end{equation}
The states in $\boldsymbol{B}_{it}$ evolve according to a random walk process,
\begin{equation}
\boldsymbol{B}_{it} = \boldsymbol{B}_{it-1}+\boldsymbol{v}_t,~\text{ with } \boldsymbol{v}_t \sim \mathcal{N}(\boldsymbol{0}, \boldsymbol{\Omega}_i), \label{eq: state_i}
\end{equation}
where $\boldsymbol{\Omega}_i = \text{diag}(\omega_1, \dots, \omega_{K_i})$ is a diagonal variance-covariance matrix. Note that if a given diagonal element of $\boldsymbol{\Omega}_i$ is zero, the corresponding regression coefficient is assumed to be constant over time. 

Typically, conjugate inverted Gamma priors are specified on $\omega_j$ ($j=1,\dots,K_i$). However, as \cite{fruhwirth2010stochastic} demonstrate, this choice is suboptimal if $\omega_j$ equals zero, since the inverted Gamma distribution artificially places prior mass away from zero and thus introduces time variation even if the likelihood points toward a constant parameter specification. To alleviate such concerns, \cite{fruhwirth2010stochastic} exploit the non-centered parameterization of Eqs.\ (\ref{eq: regression_i}) and (\ref{eq: state_i}),
\begin{equation}
y_{it} = \boldsymbol{B}'_{i0} \boldsymbol{z}_{it}+\tilde{\boldsymbol{B}}'_{it} \sqrt{\boldsymbol{\Omega}_i} \boldsymbol{z}_{it} + \eta_{it}. \label{eq: regression_NC}
\end{equation}
We let $\sqrt{\boldsymbol{\Omega}_i}$ denote the matrix square root such that $\boldsymbol{\Omega}_i=\sqrt{\boldsymbol{\Omega}_i}\sqrt{\boldsymbol{\Omega}_i}$ and $\tilde{\boldsymbol{B}}_{it}$ has typical element $j$ given by $\tilde{b}_{ij,t}=\frac{ b_{ij,t}-b_{ij,0}}{\sqrt{\omega_{ij}}}$. The corresponding state equation is given by
\begin{equation}
\tilde{\boldsymbol{B}}_{it} = \tilde{\boldsymbol{B}}_{it-1}+\boldsymbol{u}_{it},~\text{ with } \boldsymbol{u}_{it} \sim \mathcal{N}(\boldsymbol{0}, \boldsymbol{I}_{K_i}).
\end{equation}
Moving from the centered to the non-centered parameterization allows us to treat the (signed) square root of the state innovation variances as additional regression parameters to be estimated. Moreover, this parameterization also enables us to control for model uncertainty associated with whether a given element of $\boldsymbol{z}_{it}$, i.e.,\ both autoregressive coefficients and covariance parameters, should be included or excluded from the model. This can be achieved by noting that if $b_{ij, 0} \neq 0$, the $j$th regressor is included. The second dimension of model uncertainty stems from the empirically relevant question whether a given regression coefficient should be constant or time-varying. Thus, if $\omega_{jj} \neq 0$, the $j$th regressor drifts smoothly over time. Especially for forecasting applications, appropriately selecting which subset of regression coefficients should be constant or time-varying proves to be one of the key determinants in achieving superior forecasting properties  \citep{d2013macroeconomic, korobilis2013hierarchical, belmonte2014hierarchical, bitto2015achieving}.

Finally, we also have to introduce a suitable law of motion for the diagonal elements of $\boldsymbol{V}_t$. Here, we assume that each $v_{it}$ evolves according to an independent AR(1) process,
\begin{equation}
v_{it}=\mu_i + \rho_i (v_{it-1}-\mu_i) + w_{it}, \quad  w_{it} \sim \mathcal{N}(0,\sigma_i^2), \label{eq: stateLOGVOLA}
\end{equation}
for $i=1,\dots, M$. The parameter $\mu_i$ denotes the mean of the $i$th log variance, $\rho_i$ is the corresponding persistence parameter, and $\sigma_i^2$ stands for the error variance of the relevant shocks.

\section{Prior specification}
We opt for a fully Bayesian approach to estimation, inference, and prediction. This calls for the specification of suitable priors on the parameters of the model. Typically, inverse Gamma or inverted Wishart priors are used for the state innovation variances in \autoref{eq: state_i}. However, as discussed above,
such priors bound the diagonal elements of $\boldsymbol{\Omega}_i$ artificially away from zero, always inducing at least some movement in the parameters of the model.

We proceed by utilizing two flexible global-local (GL) shrinkage priors \citep[see][]{polson2010shrink} on $\boldsymbol{B}_{i0}$ and $\boldsymbol{\omega}_i = (\omega_{i1}, \dots, \omega_{i K_i})'$. A GL shrinkage prior comprises a global scaling parameter that pushes all elements of the coefficient vector toward zero and a set of local scaling parameters that enable coefficient-specific deviations from this general pattern. 

\subsection{The NG shrinkage prior}
The first prior we consider is a modified variant of the NG shrinkage prior proposed in \cite{griffin2010inference} and adopted within the general class of state space models in \cite{bitto2015achieving}. In what follows, we let $\boldsymbol{a}_{0}=\text{vec}(\boldsymbol{A}_{0})$ denote the time-invariant part of the VAR coefficients  with typical element $a_{0j}$ for $j=1,\dots,K=pM^2$. The corresponding signed squared root of the state innovation variance is consequently denoted by $\pm \sqrt{\omega}_{j}$ or simply $\sqrt{\omega}_{j}$.  Thus, $\sqrt{\omega}_{j}$ crucially determines the amount of time variation in the $j$th element of $\boldsymbol{a}_t$. 

 With this in mind, our prior specification is a scale mixture of Gaussians,
\begin{align}
a_{0j}| \tau_{a j}^2, \lambda_a &\sim \mathcal{N}(0, 2/\lambda_a ~ \tau_{a j}^2),  \quad \tau_{a j}^2 \sim \mathcal{G}(\vartheta_a, \vartheta_a)\\
\sqrt{\omega}_{j}| \tau_{\omega j}^2, \lambda_{\omega} &\sim \mathcal{N}(0, 2/\lambda_{\omega} ~ \tau_{\omega j}^2), \quad \tau_{\omega j}^2 \sim \mathcal{G}(\vartheta_\omega,\vartheta_\omega)\\
\lambda_a &\sim \mathcal{G}(c_{\lambda_a}, d_{\lambda_a}), \\
\lambda_\omega &\sim \mathcal{G}(c_{\lambda_\omega}, d_{\lambda_\omega}),
\end{align}

where $\tau_{s j}^2$ for $s \in \{a, \omega\}$ is  a  set of  local scaling parameters that follow a  Gamma distribution and $\lambda_s$ is a set of global shrinkage parameters. 
The hyperparameter  $\vartheta_a$ controls the excess kurtosis of the marginal prior,
\begin{equation}
p(a_{0j}|\lambda_a) = \int p(a_{0j}|\tau_{a j}^2 ,\lambda_a) d \tau_{a j}^2,
\end{equation}
obtained after integrating out the local scaling parameters. For the marginal prior, $\lambda_a$ controls the overall degree of shrinkage. Lower values of $\vartheta_a$ place increasing prior mass at zero while at the same time lead to heavy tails of $p(a_{0j}|\lambda_a)$. 

On the covariance parameters $\tilde{h}_{is,0}~(i=2,\dots, M; s=pM+1,\dots, K_i)$ and the associated innovation standard deviations $\gamma_{is} = \sqrt{\omega}_{i s}$, we impose the standard implementation of the NG prior. To simplify prior implementation we collect the $v=M (M-1)/2$ free covariance parameters in a vector $\tilde{\boldsymbol{h}}_0$ and the corresponding elements of $\boldsymbol{\Omega}=\text{diag}(\boldsymbol{\Omega}_1, \dots, \boldsymbol{\Omega}_M)$ in a $v$-dimensional vector $\boldsymbol{\gamma}$ with typical elements $\tilde{h}_{i0}$ and $\gamma_{i}$,
\begin{align}
\tilde{h}_{i0}|\tau_{h i}^2, \varpi_h &\sim \mathcal{N}(0, 2/\varpi_h ~ \tau_{h i}^2), \quad \tau_{h  i}^2 \sim \mathcal{G}(\vartheta_{h},\vartheta_{h}),\\
\gamma_{i}| \tau^2_{\gamma i}, \varpi_\gamma &\sim \mathcal{N}(0, 2/\varpi_\gamma ~ \tau^2_{\gamma i}), \quad \tau^2_{\gamma i} \sim \mathcal{G}(\vartheta_{\gamma},\vartheta_{\gamma}),\\
 \varpi_h &\sim \mathcal{G}(c_{\varpi_h}, d_{\varpi_h}), \\
 \varpi_\gamma &\sim \mathcal{G}(c_{\varpi_\gamma}, d_{\varpi_\gamma}).
\end{align}
Here, for $s \in \{h, \gamma\}$, $\tau_{s i}^2$ is the set of local scaling parameters, and $\varpi_h$ and $\varpi_\gamma$ are global shrinkage parameters that push all covariance parameters and the corresponding state innovation standard deviations across equations to zero, respectively.  The set of hyperparameters $\vartheta_s$ for $s \in \{h, \gamma\}$ again controls the excess kurtosis of the marginal priors.  

\subsection{The DL shrinkage prior}
The NG prior possesses good empirical properties. However, from a theoretical point of view, its properties are still not well understood. In principle, GL shrinkage priors aim to approximate a standard spike and slab prior \citep{george1993variable, george2008bayesian} by introducing suitable mixing distributions on the local and global scaling parameters of the model.  \cite{bhattacharya2015dirichlet} introduce a prior specification and analyze its properties within the stylized normal means problem. Their prior, the DL shrinkage prior, excels in both theory and empirical applications, especially in very high dimensions. Thus, for the TVP-VAR-SV, it seems to be well suited given the large dimensional parameter and state space.

Similarly to the NG prior, the DL prior also depends on a set of global and local shrinkage parameters,
\begin{align}
a_{0j}| \psi_{a j}, \xi^2_{a j},   \tilde{\lambda}^2_a &\sim \mathcal{N}(0, \psi_{a j} \xi^2_{a j} \tilde{\lambda}_a^2),  \quad \psi_{a j} \sim {Exp}(1/2), \quad \xi_{j} \sim Dir(n_a,\dots,n_a),\\
\sqrt{\omega}_{j}| \psi_{\omega j}, \xi^2_{\omega j}, \tilde{\lambda}^2_\omega &\sim \mathcal{N}(0, \psi_{\omega j} \xi^2_{\omega j} \tilde{\lambda}_\omega^{2}),  \quad \psi_{\omega j} \sim {Exp}(1/2), \quad \xi_{j} \sim Dir(n_\omega,\dots,n_\omega),\\
\tilde{\lambda}_a &\sim \mathcal{G}(K n_a, 1/2),\\
\tilde{\lambda}_\omega &\sim \mathcal{G}(K n_\omega, 1/2). \label{eq: globalpriorDL}
\end{align}
Hereby, for $s\in\{a, \omega\}$, $\xi_{s j}$ is the set of local scaling parameters defined on the $(K-1)$-dimensional unit simplex $\mathcal{S}^{K-1}=\{\boldsymbol{x}=(x_1,\dots,x_K)': {x}_{j} \ge 0, \sum_{j=1}^{K} {x}_{j}=1 \}$ with $\boldsymbol{\xi}_s=(\xi_{s1},\dots, \xi_{sK})'$. The parameter $n_s$ controls the overall tightness of the prior. $\psi_{sj}$ is the set of auxiliary scaling parameters to achieve conditional normality.

For the variance-covariance matrix we also impose the DL prior,
 \begin{align}
\tilde{h}_{i 0}|\psi_{h i}, \xi^2_{h i}, \tilde{\varpi}^2_h &\sim \mathcal{N}(0, \psi_{h i}  \xi^2_{h i} \tilde{\varpi}_h^{2}), \quad \psi_{h i}^2 \sim {Exp}(1/2), \quad \xi_{h i} \sim Dir(n_h, \dots, n_h),\\
\gamma_{i}| \psi_{\gamma i}, \xi^2_{\gamma i}, \tilde{\varpi}^2_\gamma &\sim \mathcal{N}(0, \psi_{\gamma i}  \xi^2_{\gamma i} \tilde{\varpi}_\gamma^{2}), \quad \psi^2_{\gamma i} \sim {Exp}(1/2), \quad \xi_{\gamma i} \sim Dir(n_\gamma, \dots, n_\gamma),\\
\tilde{\varpi}_h &\sim \mathcal{G}( v n_h, 1/2), \\
\tilde{\varpi}_\gamma &\sim \mathcal{G}( v n_\gamma, 1/2). 
\end{align}
The local shrinkage parameters $\psi_{si}$ and $\xi^2_{s i}$ for $s \in \{h, \gamma\}$ are defined analogously to the case of the regression coefficients described above. We let $\tilde{\varpi}_s$ denote the set of global shrinkage parameters with large values, implying heavy shrinkage on the covariance parameters of the model.

The main differences of the NG and the DL prior are the presence of the Dirichlet components that introduce even more flexibility. \cite{bhattacharya2015dirichlet} show that in the framework of the stylized normal means problem, this specification yields excellent posterior contraction rates in light of a sparse DGP. Within an extensive simulation exercise, they moreover provide some evidence that this prior also works well in practice.

Finally, the prior setup on the coefficients in the state equation of the log volatilities closely follows \cite{kastner2016dealing}. Specifically, we place a weakly informative Gaussian prior on $\mu_i$, $\mu_i \sim \mathcal{N}(0,10^2)$ and a Beta prior on $\frac{\rho_i +1}{2} \sim \mathcal{B}(25, 1.5)$. Additionally, $\sigma_i^2 \sim \mathcal{G}(1/2, 1/2)$ introduces some shrinkage on the process innovation variances of the log volatilities. This setup is used for all equations.

\section{Bayesian inference}
The joint posterior distribution of our model is analytically intractable. Fortunately, however, the full conditional posterior distributions mostly belong to some well-known family of distributions, implying that we can set up a conceptually straightforward Gibbs sampling algorithm to estimate the model.

\subsection{A brief sketch of the MCMC algorithm}

Our algorithm is related to the MCMC scheme put forward in \cite{carriero2016large} and estimates the latent states on an equation-by-equation basis. Specifically, conditional on a suitable set of initial conditions, the algorithm cycles through the following steps:
\begin{enumerate}
 \item Draw $(\boldsymbol{B}_{i0}', \omega_{i1},\dots, \omega_{i K_i})'$ for $i=1,\dots,M$ from $\mathcal{N}(\boldsymbol{\mu}_{B i}, \boldsymbol{V}_i)$ with  $\boldsymbol{V}_i= (\boldsymbol{Z}'_i\boldsymbol{Z}_i+ \underline{\boldsymbol{V}}_i^{-1})^{-1}$ and $\boldsymbol{\mu}_{B i}=\boldsymbol{V}_i ( \boldsymbol{Z}_i \boldsymbol{Y}_i)$. We let $\boldsymbol{Z}_i$ be a $T \times (2 K_i)$ matrix with typical $t$th row $[\boldsymbol{z}'_{it}, (\boldsymbol{B}_{it}  \odot\boldsymbol{z}_{it})']~e^{-(v_{it}/2)}$, $\boldsymbol{Y}_i$ is a $T$-dimensional vector with element $y_{it} ~e^{-(v_{it}/2)}$, and $\underline{\boldsymbol{V}}_i$ is a prior covariance matrix that depends on the prior specification adopted. Note that in contrast to \cite{carriero2016large} who sample the VAR parameters in ${\boldsymbol{A}}_0$ and the elements of $\tilde{\boldsymbol{H}}_0$ conditionally on each other, \hl{we follow \cite{eisenstat2016stochastic} and sample these jointly}, which speeds up the mixing of the sampler.

\item Simulate the full history of $\{\tilde{\bm{B}}_{it}\}_{t=1}^T$ by means of a forward filtering backward sampling algorithm \citep[see][]{carter1994gibbs, fruhwirth1994data} per equation. 

\item The log volatilities and the corresponding parameters of the state equation in \autoref{eq: stateLOGVOLA} are simulated using the algorithm put forward in \cite{kastner2014ancillarity} via the R package stochvol \citep{kastner2016dealing, stochvol2}.

\item Depending on the prior specification adopted, draw the parameters used to construct $\underline{\boldsymbol{V}}_i$ using the conditional posterior distributions detailed in Section~\ref{sec:condpostNG} (NG prior) or Section \ref{sec:condpostDL} (DL prior).
\end{enumerate}
This algorithm produces draws from the joint posterior distribution of the states and the model parameters. In the empirical application that follows we use 30,000 iterations where we discard the first 15,000 as burn-in.


\subsection{Conditional posterior distributions associated with the NG prior} \label{sec:condpostNG}
Conditional on the full history of all latent states in our model as well as the global shrinkage parameters, it is straightforward to show that the conditional posterior distributions of $\tau_{sj}^2$ for $s \in \{a, \omega\}$ and $j=1,\dots,K$ are given by 
\begin{align}
\tau_{aj}^2|\bullet \sim \mathcal{GIG}(\vartheta_{a}-1/2, a_{0j}^2, \vartheta_{a} \lambda_a), \quad 
\tau_{\omega j}^2|\bullet \sim \mathcal{GIG}(\vartheta_{\omega}-1/2, \sqrt{\omega}_{j}^2, \vartheta_{\omega} \lambda_\omega),
\end{align}
where $\bullet$ indicates conditioning on the remaining parameters and states of the model. Moreover, $\mathcal{GIG}(\zeta, \chi, \varrho)$ denotes the generalized inverse Gaussian distribution with density proportional to $x^{\zeta-1} \exp\{-(\chi/x+\varrho x)/2\}$. To draw from this distribution, we use the algorithm of \cite{hoe-ley:gen} implemented in the R package GIGrvg \citep{r:gig}.
 
The conditional posteriors of the local scalings for the covariance parameters and their corresponding innovation standard deviations also follow GIG distributions,
\begin{align}
\tau_{h i}^2|\bullet \sim \mathcal{GIG}(\vartheta_h-1/2, \tilde{h}_{i0}^2, \vartheta_h \varpi_h), \quad
\tau_{\gamma i}^2|\bullet \sim \mathcal{GIG}(\vartheta_\gamma-1/2, \gamma_{i}^2, \vartheta_\gamma \varpi_\gamma).
\end{align} 

The conditional posterior of $\lambda_s$ for $s \in \{a, \omega\}$ is of well-known form, namely, a  Gamma distribution,
\begin{equation}
\lambda_s |\bullet \sim \mathcal{G}\left\lbrace c_{\lambda_s}+ \vartheta_{s}  K, d_{\lambda_s}+\frac{\vartheta_{s}}{2} \sum_{j=1}^K \tau^2_{sj}\right\rbrace.
\end{equation}
Likewise, for $s \in \{h,\gamma\}$, the conditional posterior of $\varpi_s$ is given by
\begin{equation}
 \varpi_s|\bullet \sim \mathcal{G}\left\{c_{\varpi_s}+ \vartheta_s v, d_{\varpi_s} + \frac{\vartheta_s}{2} \sum_{i=1}^v \tau^2_{sj} \right\}.
\end{equation}
\subsection{Conditional posterior distributions associated with the DL prior}\label{sec:condpostDL}
Following \cite{bhattacharya2015dirichlet} we update jointly the full conditional posterior distribution $p(\xi_{sj}, \lambda_s, \psi_{sj} | \bullet)$ for $s \in \{a, \omega\}$. For the Dirichlet components, the conditional posterior distribution is obtained by  sampling a set of $K$ auxiliary variables $N_{aj}, N_{\omega j}~(j=1,\dots,K)$,
\begin{align}
N_{aj}|a_{0j} &\sim \mathcal{GIG}(n_a-1, 2 |a_{0j}|,1),\quad N_{\omega j}|\sqrt{\omega}_{j} \sim \mathcal{GIG}(n_\omega-1, 2 |\sqrt{\omega}_{j}|,1).
\label{eq:Naj}
\end{align}
After obtaining the $K$ scaling parameters we set  $\xi_{aj}= N_{aj}/ N_a$  and $\xi_{\omega j}= N_{\omega j}/N_\omega$ with $N_a=\sum_{j=1}^K N_{aj}$ and $ N_\omega=\sum_{j=1}^K N_{\omega j}$.

The conditional posterior of the global shrinkage parameter $\tilde{\lambda}_s$ for $s \in \{a, \omega\}$ follows a GIG distribution,
\begin{gather}
\begin{aligned}
    \tilde{\lambda}_a | \xi_{aj}, a_{0j} &\sim \mathcal{GIG}\left\{K(n_a-1), 2\sum_{j=1}^K |a_{0j}| / \xi_{aj}, 1\right\}, \\
    \tilde{\lambda}_\omega | \xi_{\omega j}, \sqrt{\omega}_j &\sim \mathcal{GIG}\left\{K(n_\omega-1), 2\sum_{j=1}^K |\sqrt{\omega}_j| / \xi_{\omega j},1\right\}.
\end{aligned}
\end{gather}
The full conditional posterior distributions of $\psi_{aj}^{-1}$ and $\psi_{\omega j}^{-1}$ are inverse Gaussian,
\begin{align}
\psi_{aj}^{-1}|\bullet \sim \mathcal{IG}(\xi_{aj} \tilde{\lambda}_a/|a_{0j}| , 1), \quad
\psi_{\omega j}^{-1}|\bullet \sim \mathcal{IG}(\xi_{\omega j} \tilde{\lambda}_\omega/|\sqrt{\omega}_j| , 1).
\end{align}

To sample from the conditional posterior distribution of $\xi_{si}$ for $s \in \{h,\gamma\}$, again, we introduce a set of auxiliary variables $N_{hi}, N_{\gamma i}$,
\begin{align}
N_{hi}|\tilde{h}_{i0} \sim \mathcal{GIG}(n_h-1, 2 |\tilde{h}_{i0}|,1),\quad N_{\gamma i}|\gamma_{i} \sim \mathcal{GIG}(n_\gamma-1, 2 |\gamma_{i}|,1),
\end{align}
and obtain draws from $\xi_{hi}$ and $\xi_{\gamma i}$ by using $\xi_{hi}= N_{hi}/\sum_{i=1}^v N_{hi}$ and $\xi_{\gamma i}= N_{\gamma i}/\sum_{i=1}^v N_{\gamma i}$.
The global shrinkage parameter on the covariance parameters and the process innovation variances follows again a GIG distribution, 
\begin{gather}
   \begin{aligned}
 \tilde{\varpi}_h | \tilde{h}_{i0}, \xi_{hi} \sim \mathcal{GIG}\left\{v (n_h-1), 2 \sum_{j=1}^v \left(\frac{|\tilde{h}_{i0}|}{\xi_{hi}} \right),1\right\}, \\
 \tilde{\varpi}_\gamma | \gamma_{i}, \xi_{\gamma i} \sim \mathcal{GIG}\left\{v (n_\gamma-1), 2 \sum_{j=1}^v \left(\frac{|\gamma_{i}|}{\xi_{\gamma i}} \right),1\right\}.
\end{aligned} 
\end{gather}
The full conditional posterior distributions of $\psi_{hi}^{-1}$ and $\psi_{\gamma i}^{-1}$ for $i=1,\dots,v$ are given by
\begin{align}
 \psi_{h i}^{-1}|\bullet &\sim \mathcal{IG}(\tilde{\varpi}_h \zeta_{h i}/  |\tilde{h}_{i0}|,1),
 \quad \psi_{\gamma i}^{-1}|\bullet \sim \mathcal{IG}(\tilde{\varpi}_\gamma \zeta_{\gamma i}/  |\gamma_{i}|,1).
\end{align}

\section{Forecasting macroeconomic quantities for three major economies}
In what follows we systematically assess the relationship between model size and model complexity by forecasting several macroeconomic indicators for three large economies, namely the EA, the UK and the US. In Section~\ref{data}, we briefly describe the different data sets and discuss model specification issues. Section~\ref{visualize} deals with simple visual summaries of posterior sparsity in terms of the VAR coefficients and their time variation for the two shrinkage priors proposed. The main forecasting results are discussed in Section~\ref{forecast}. Finally, Section~\ref{pool} discusses the possibility to dynamically select among different specifications in an automatic fashion.

\subsection{Data and model specification} \label{data}
We use prominent macroeconomic data sets for the EA, the UK and the US. All three data sets are on a quarterly frequency but span different periods of time. For the euro area, we take data from the area wide model \citep{awm} and additionally include equity prices available from 1987Q1 to 2015Q4. UK data stem from the Bank of England's ``A millennium of macroeconomic data'' \citep{ukdata} and covers the period from 1982Q2 to 2016Q4. For the US, we use a subset from the FRED QD database \citep{McCracken2016}, which covers the period from 1959Q1 to 2015Q1. 

For each of the three cases we use three subsets, a small (3 variables), a medium (7 variables) and a large (15 variables) subset. The small subset covers only real activity, prices and short-term interest rates. The medium models cover in addition investment and consumption, the unemployment rate, and either nominal or effective exchange rates. For the large models we add wages, money (measured as M2 or M3), government consumption, exports, equity prices, and 10-year government bond yields. 

To complete the data set for the large models, we include additional variables depending on data availability for each country set. For example, the UK data set offers a wide range of financial data, so we complement the large model by including also data on mortgage rates and bond spreads. For the  EA data set we include also a commodity price indicator and labor market productivity, while for the US we add consumer sentiment and hours worked. In what follows, we are interested not only in the relative performance of the different priors but also in the forecasting performance using different information sets. Thus, we have opted firstly to strike a good balance between different types of data (e.g., real, labor market, and financial market data) and secondly to alter variables for the large data sets slightly. This is done to rule out that performance between information sets depends crucially on the type of information that is added (e.g., labor market data vs.\ financial market data).  

For data that are non-stationary  we take first differences; see \autoref{tbl:data} in the appendix for more details. Consistent with the literature \citep{cogley2005drifts, primiceri2005time, d2013macroeconomic}, we include $p = 2$ lags of the endogenous variables in all models. 
For both the NG prior and the DL prior, we use lag-wise specifications, similarly to those in \cite{Huber2017}, in order to increase shrinkage for higher lags.\footnote{Given that the lag length is set equal to two, this choice does not matter much empirically.}


\subsection{Inspecting posterior sparsity} \label{visualize}

\begin{figure}[tp]
\subfigure[VAR coefficients: DL prior]{\includegraphics[width=.496\textwidth]{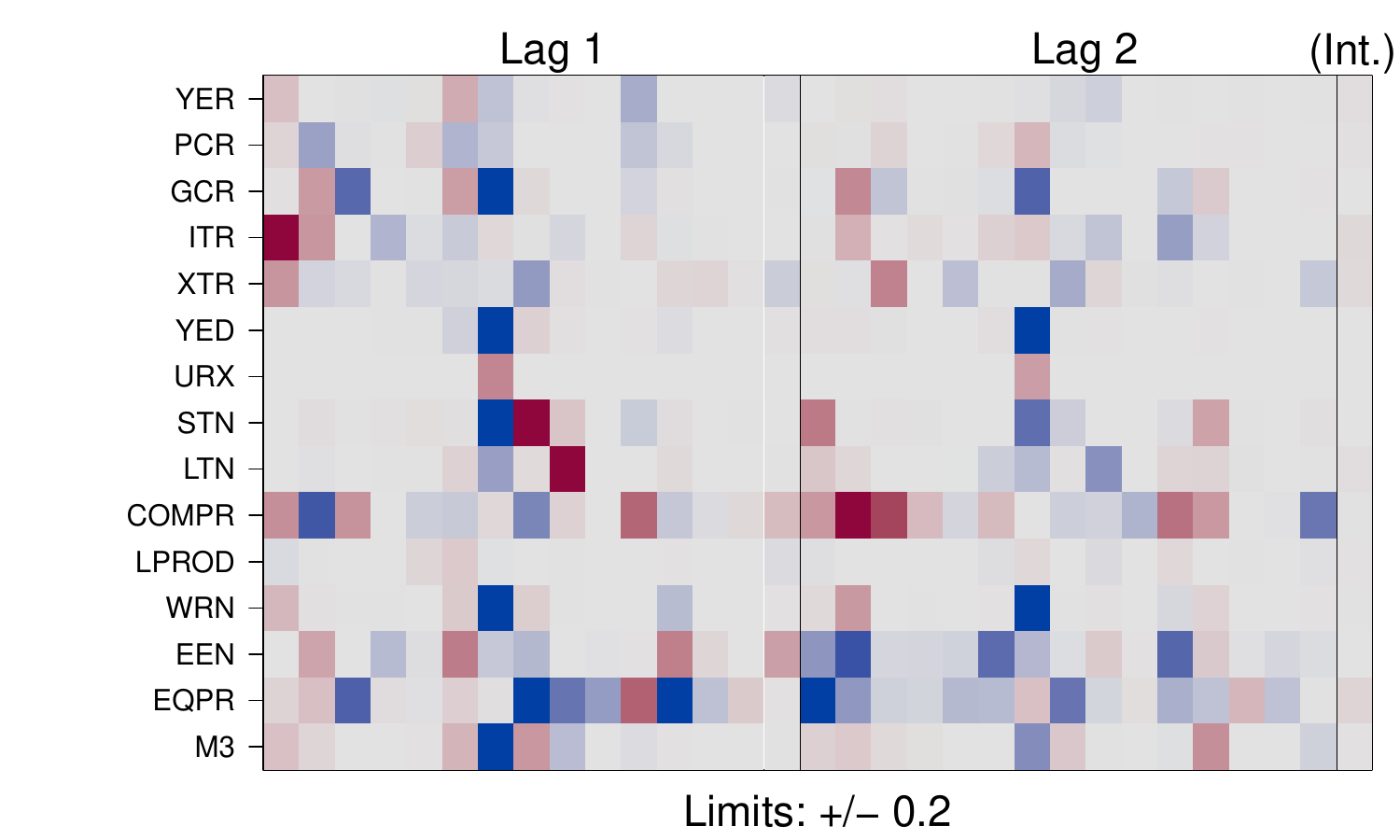}}
\subfigure[VAR coefficients: NG prior]{\includegraphics[width=.496\textwidth]{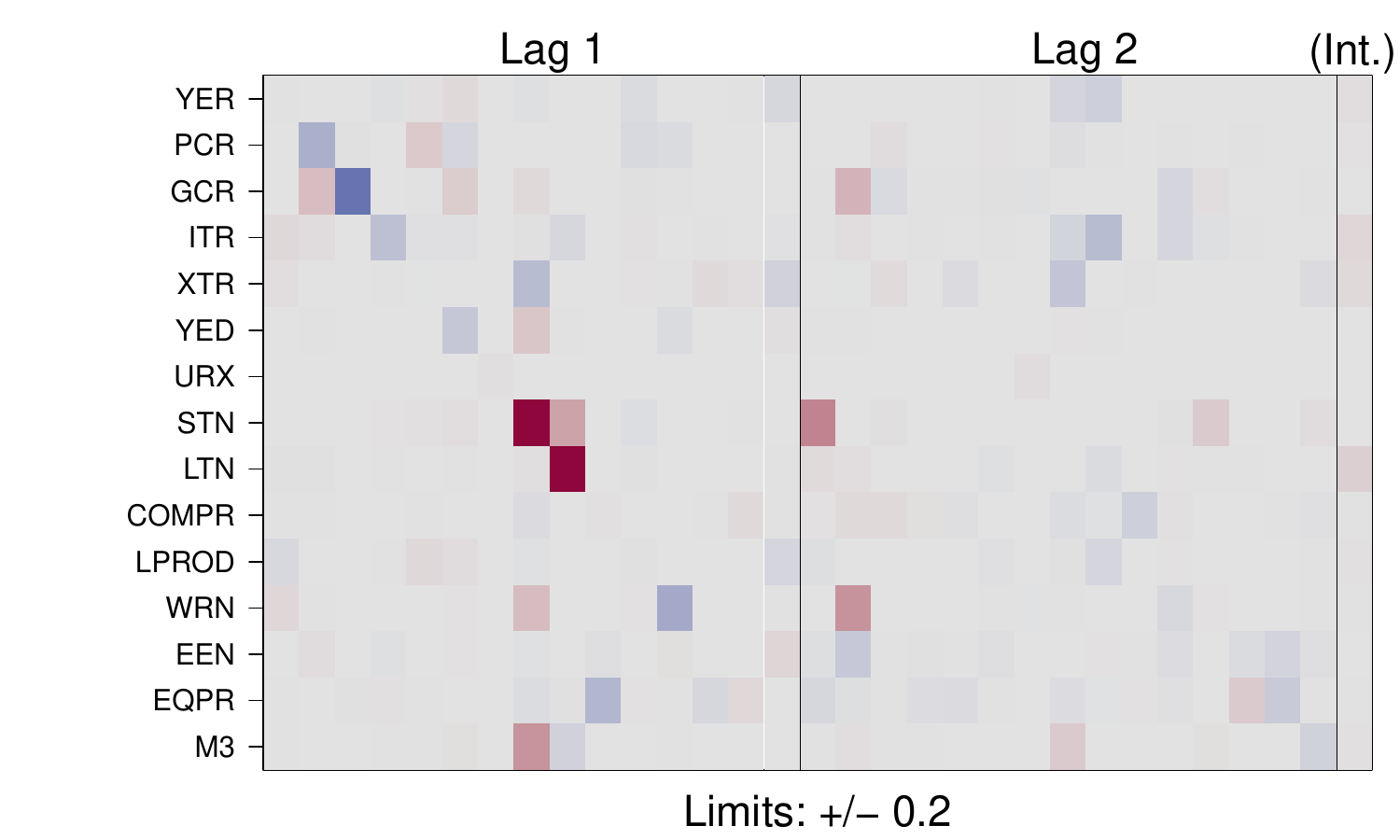}}\\
\subfigure[State innovation sds: DL prior]{\includegraphics[width=.496\textwidth]{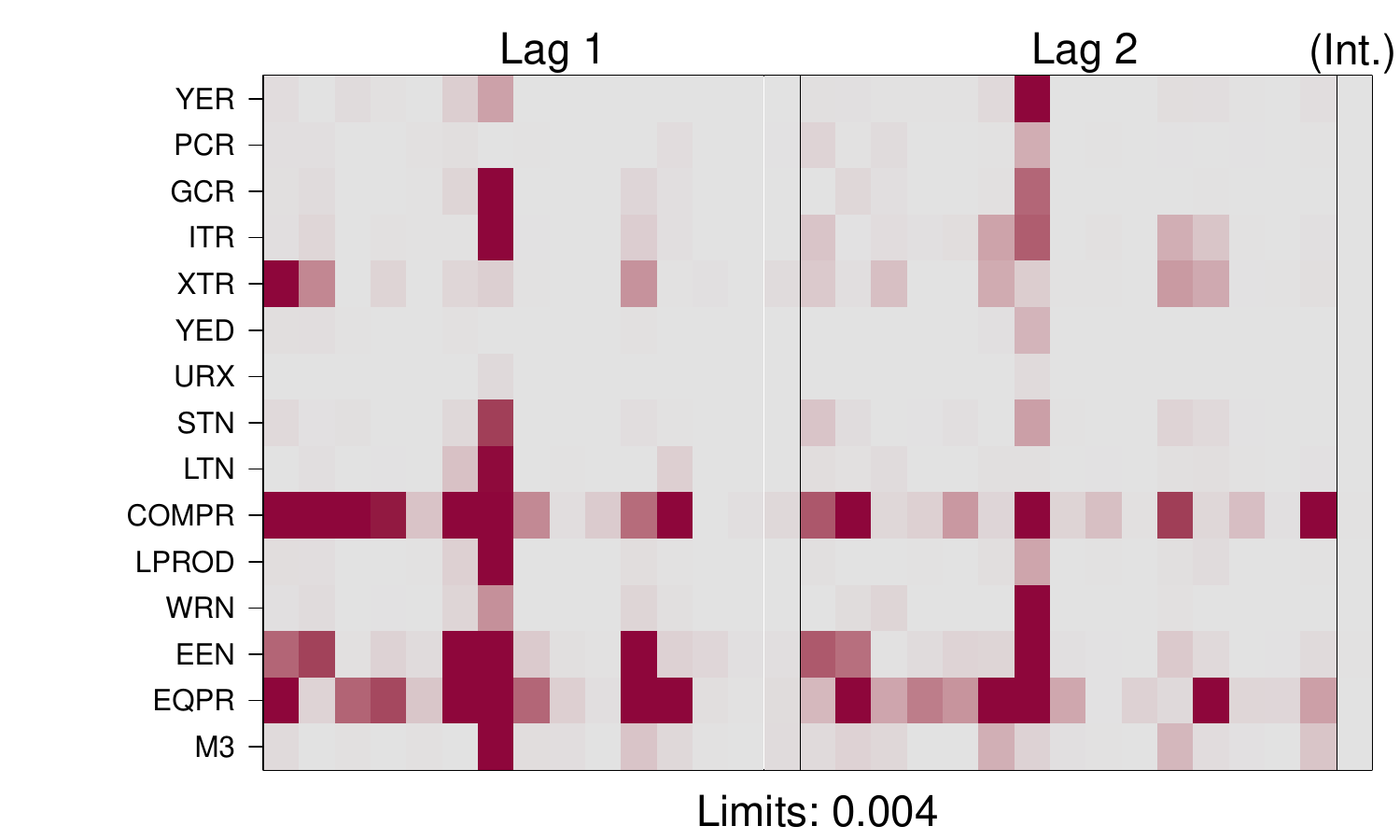}}
\subfigure[State innovation sds: NG prior]{\includegraphics[width=.496\textwidth]{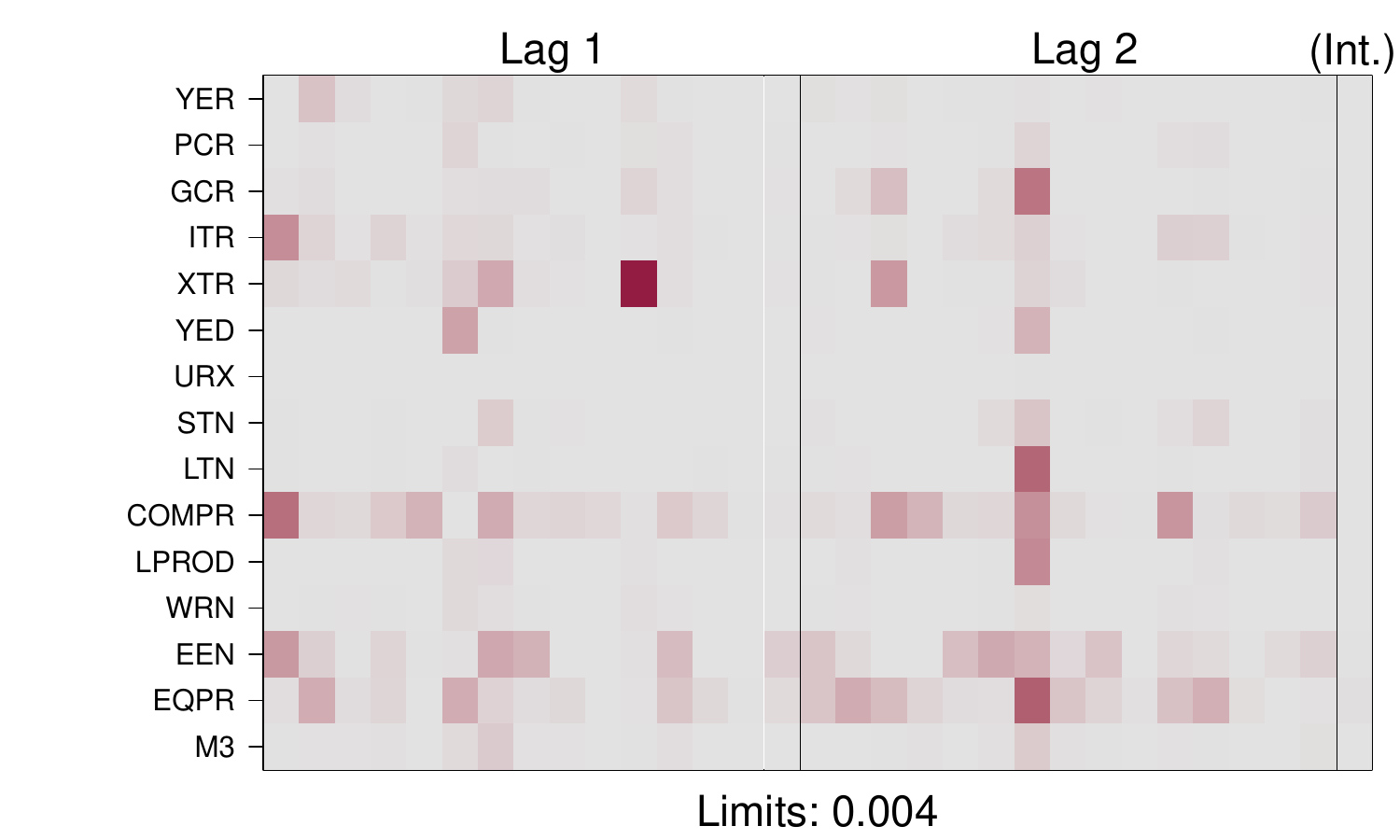}}\\
\caption{Posterior means in the large model -- euro area.} \label{fig:heat_ea}
\end{figure}

\begin{figure}[tp]
\subfigure[Posterior coefficients: DL]{\includegraphics[width=.496\textwidth]{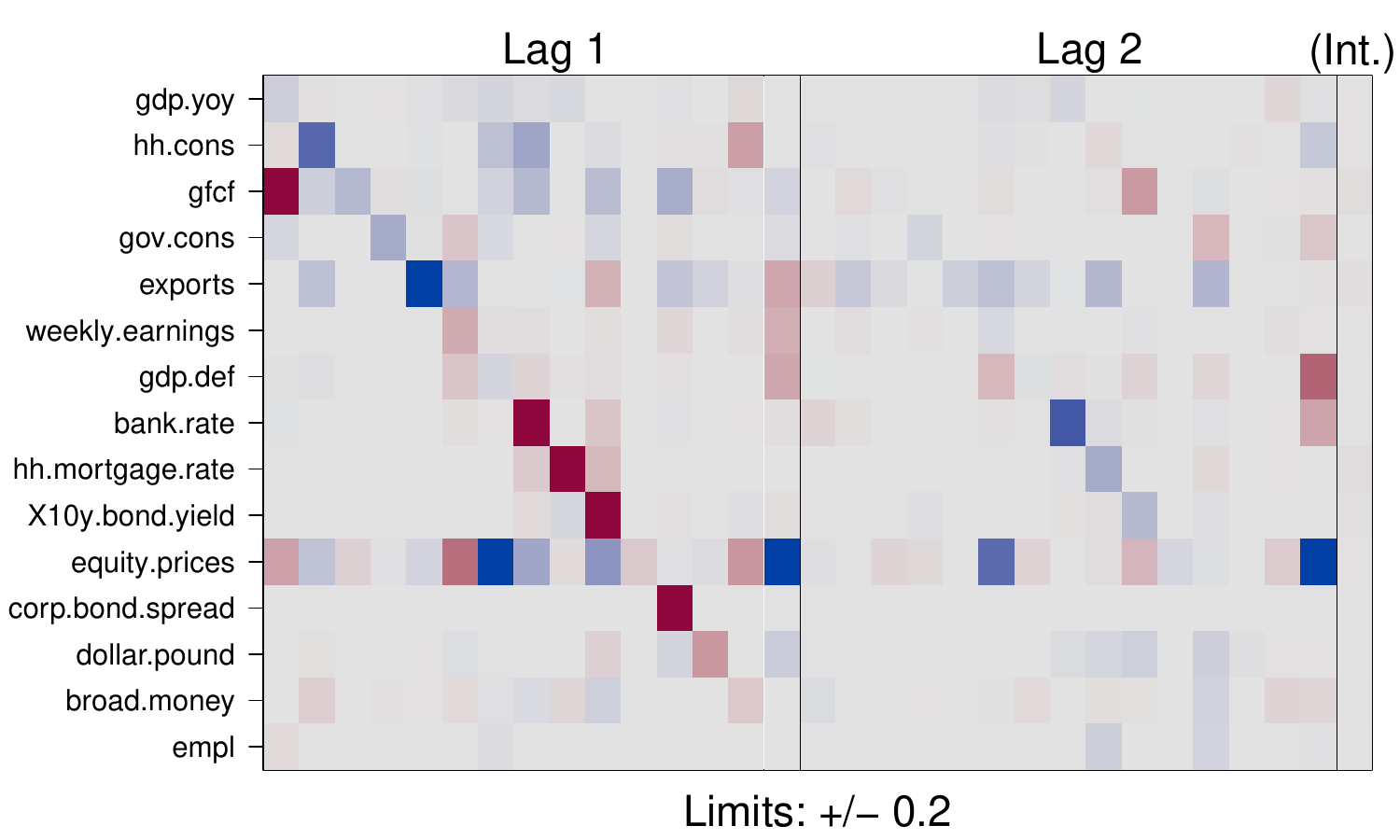}}
\subfigure[Posterior coefficients: NG]{\includegraphics[width=.496\textwidth]{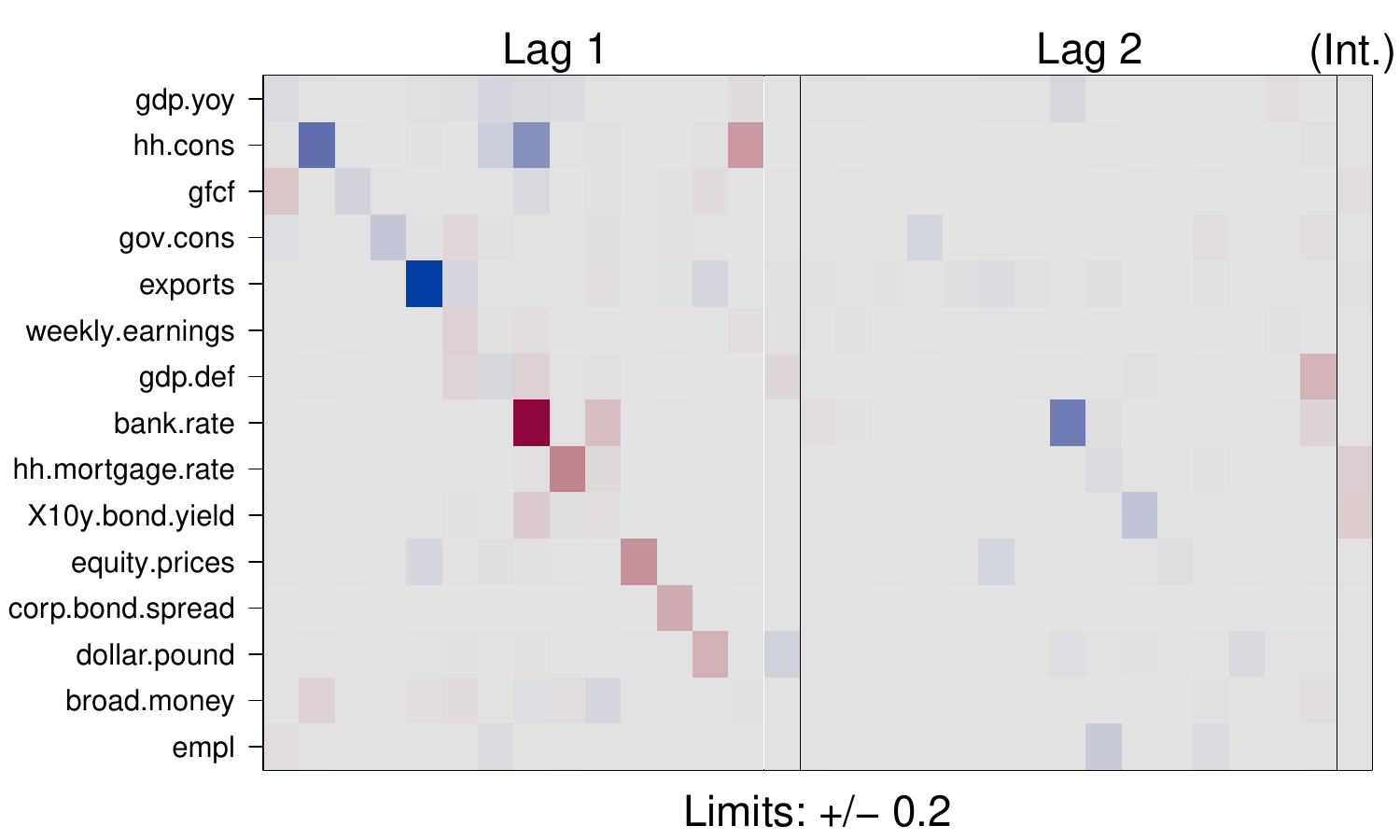}}\\
\subfigure[State innovation sds: DL]{\includegraphics[width=.496\textwidth]{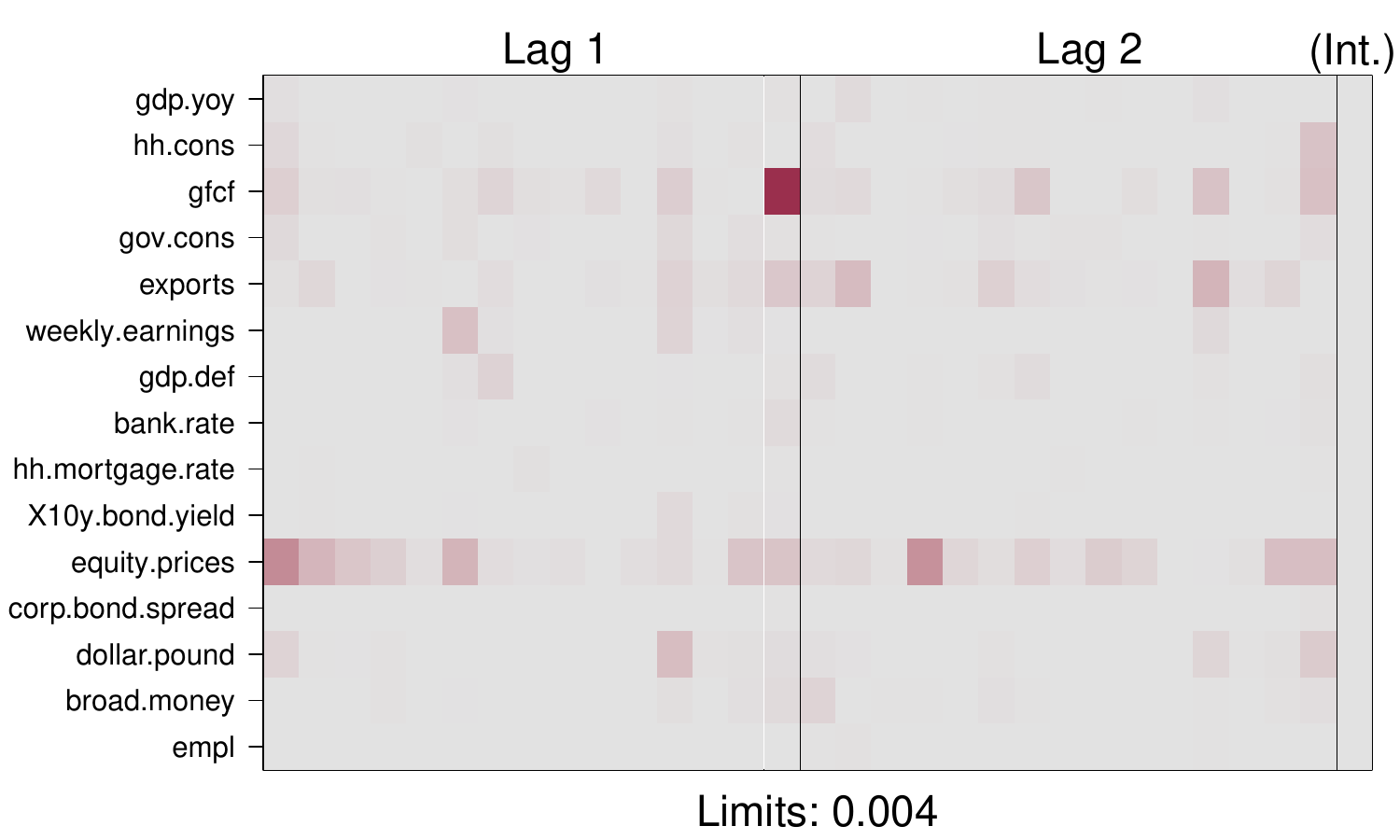}}
\subfigure[State innovation sds: NG]{\includegraphics[width=.496\textwidth]{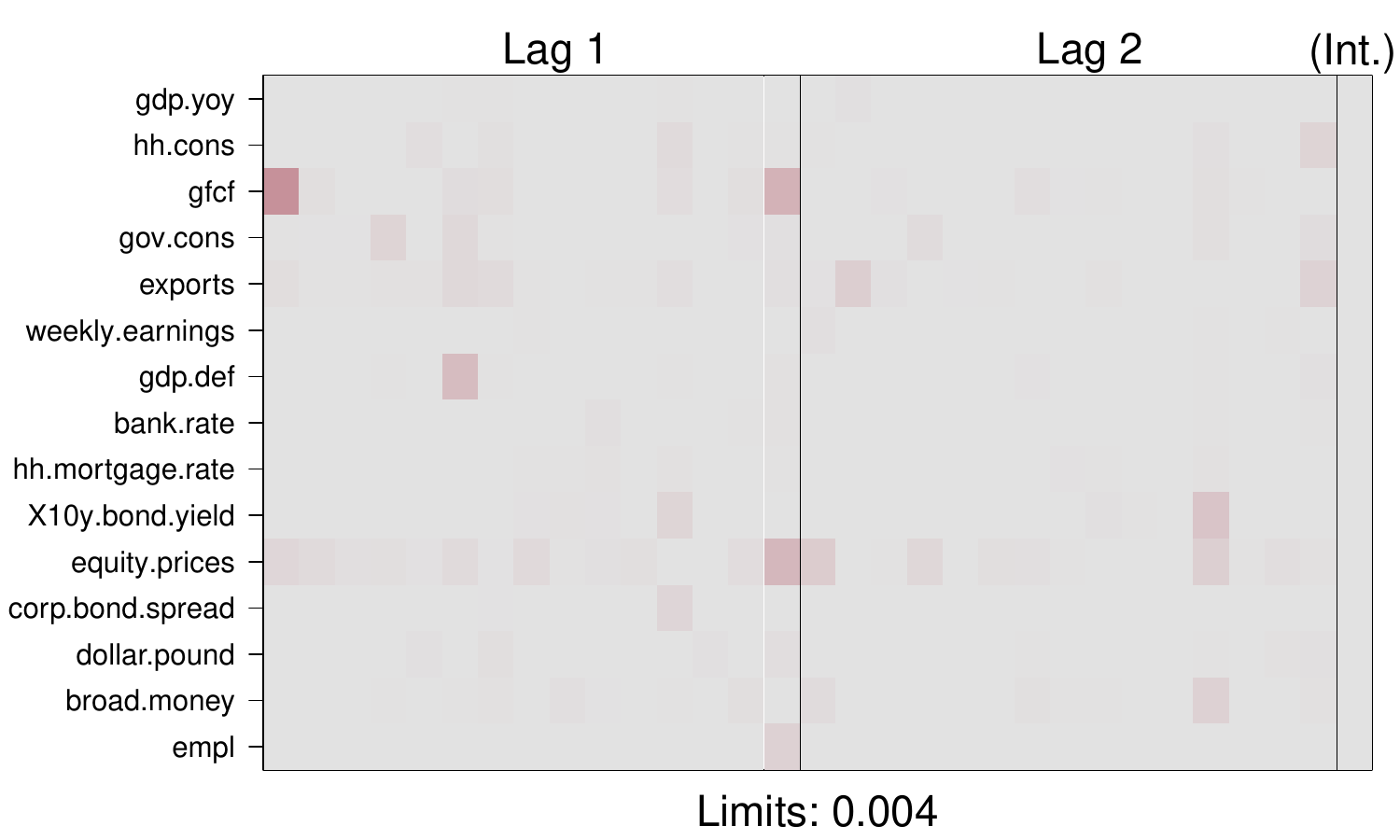}}\\
\caption{Posterior means in the large model -- UK.} \label{fig:heat_uk}
\end{figure}

\begin{figure}[tp]
\subfigure[Posterior coefficients: DL]{\includegraphics[width=.496\textwidth]{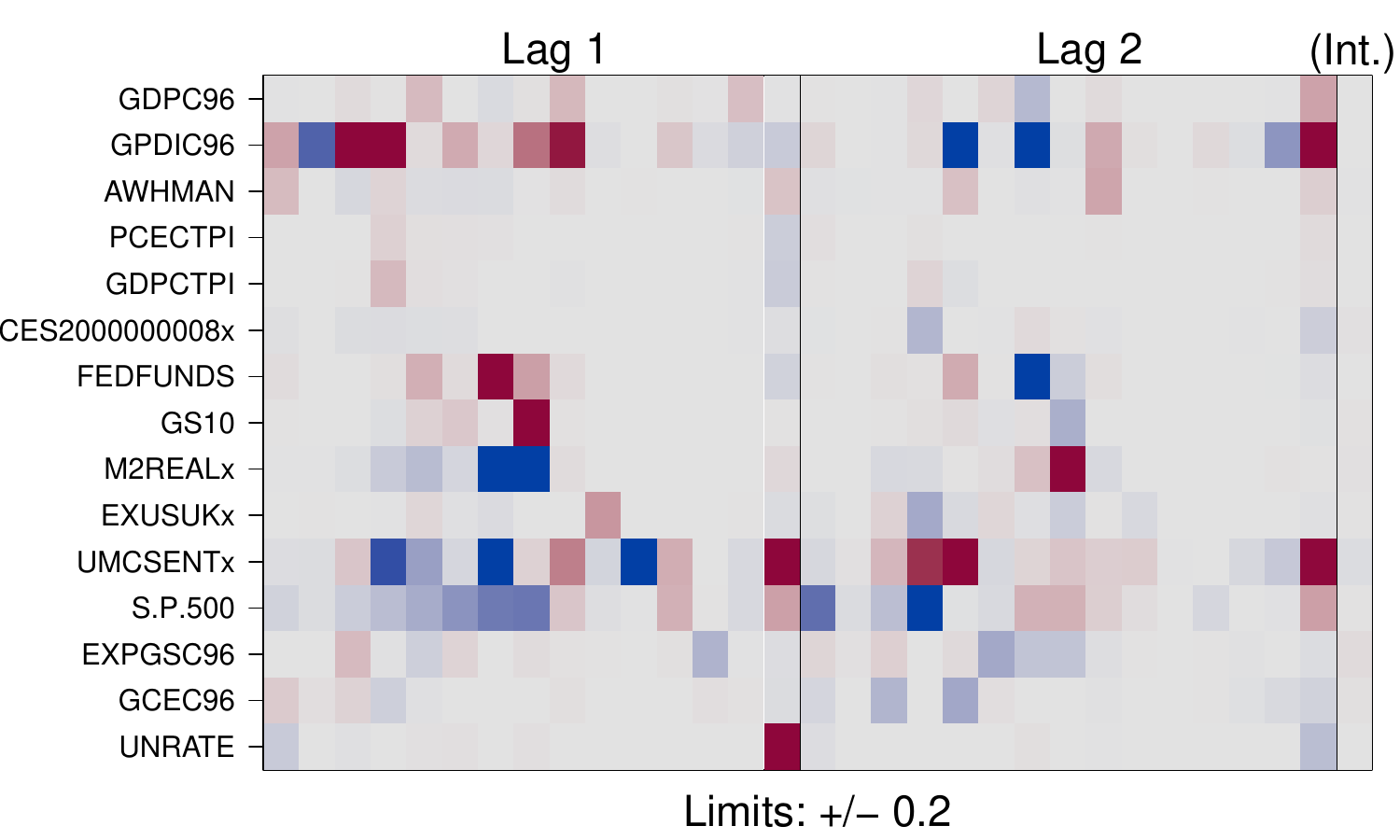}}
\subfigure[Posterior coefficients: NG]{\includegraphics[width=.496\textwidth]{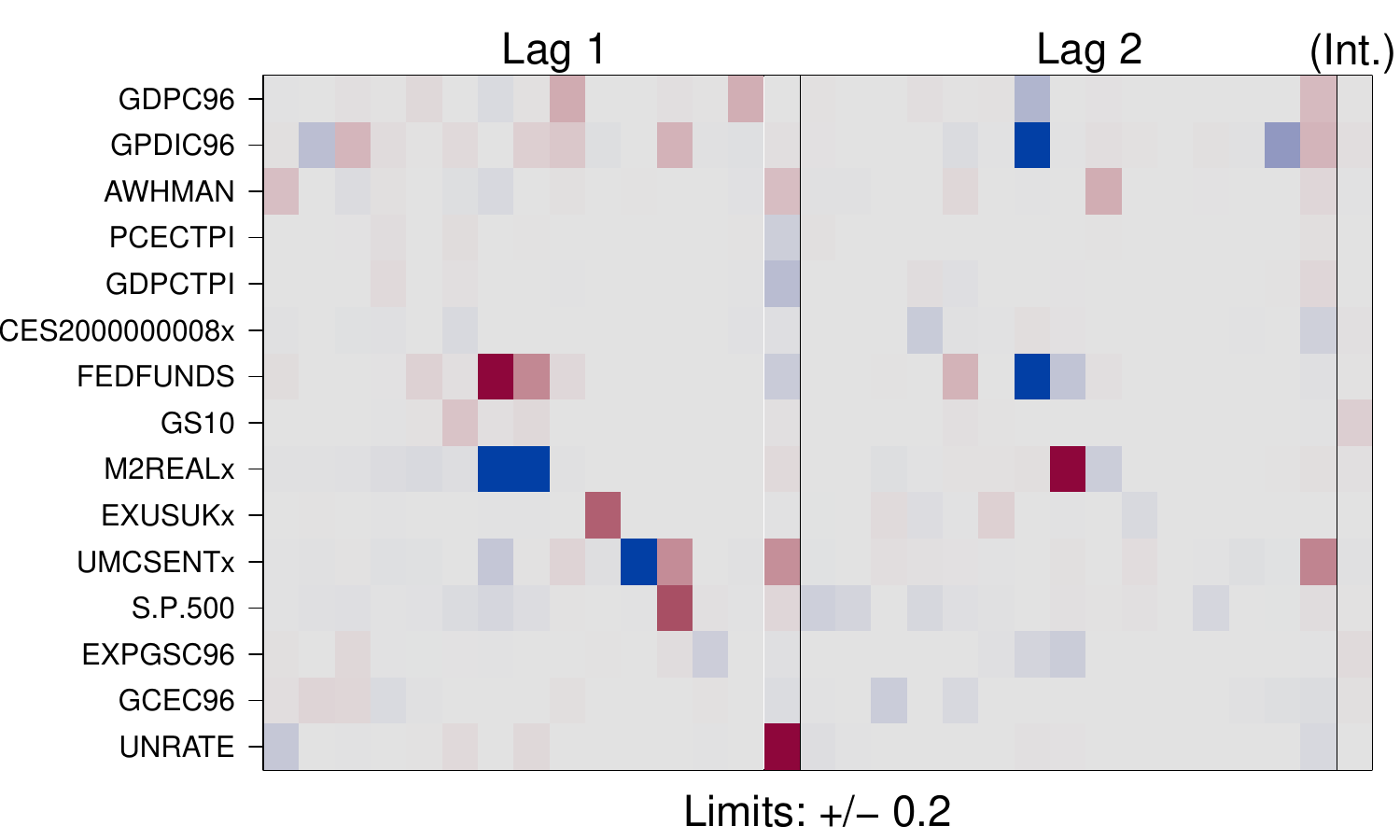}}\\
\subfigure[State innovation sds: DL]{\includegraphics[width=.496\textwidth]{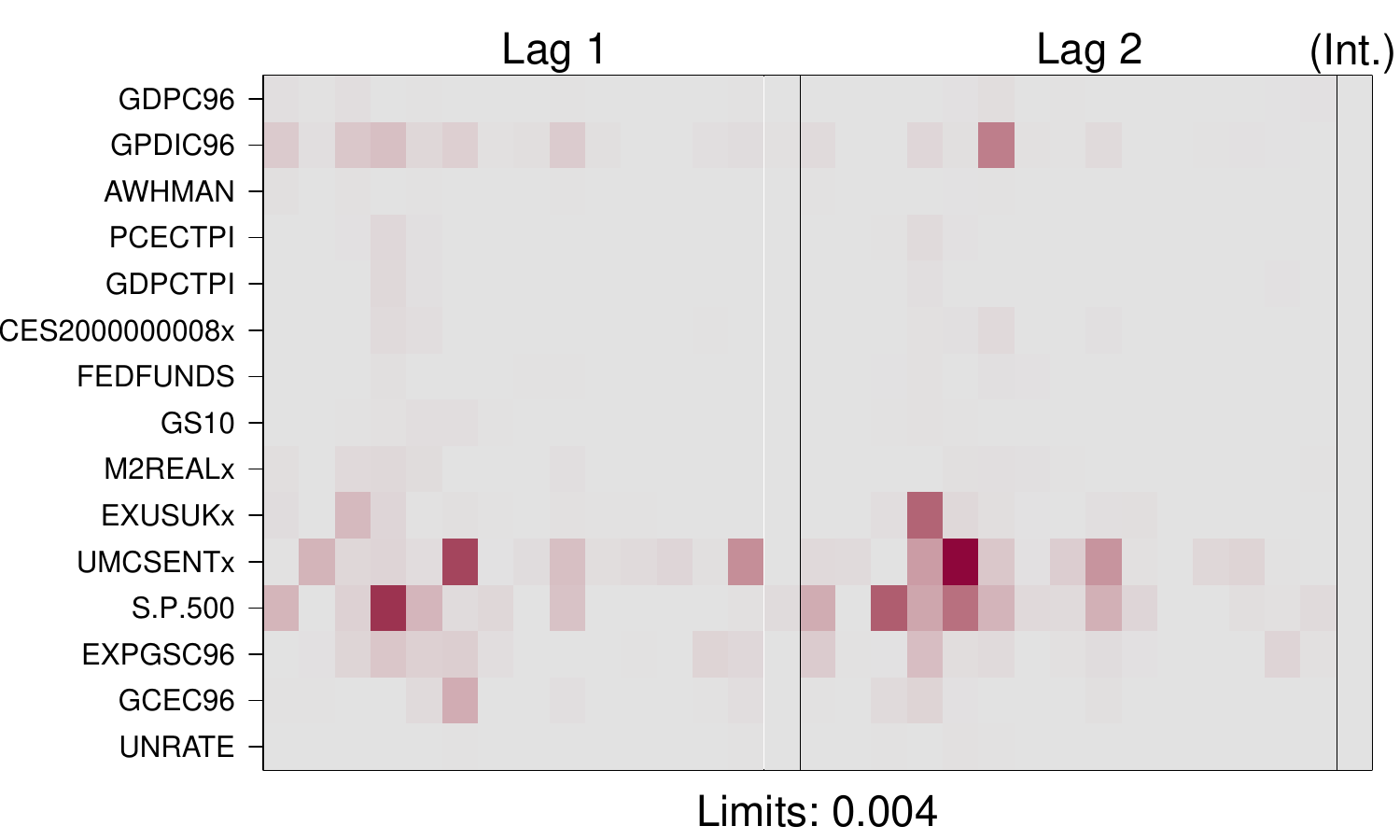}}
\subfigure[State innovation sds: NG]{\includegraphics[width=.496\textwidth]{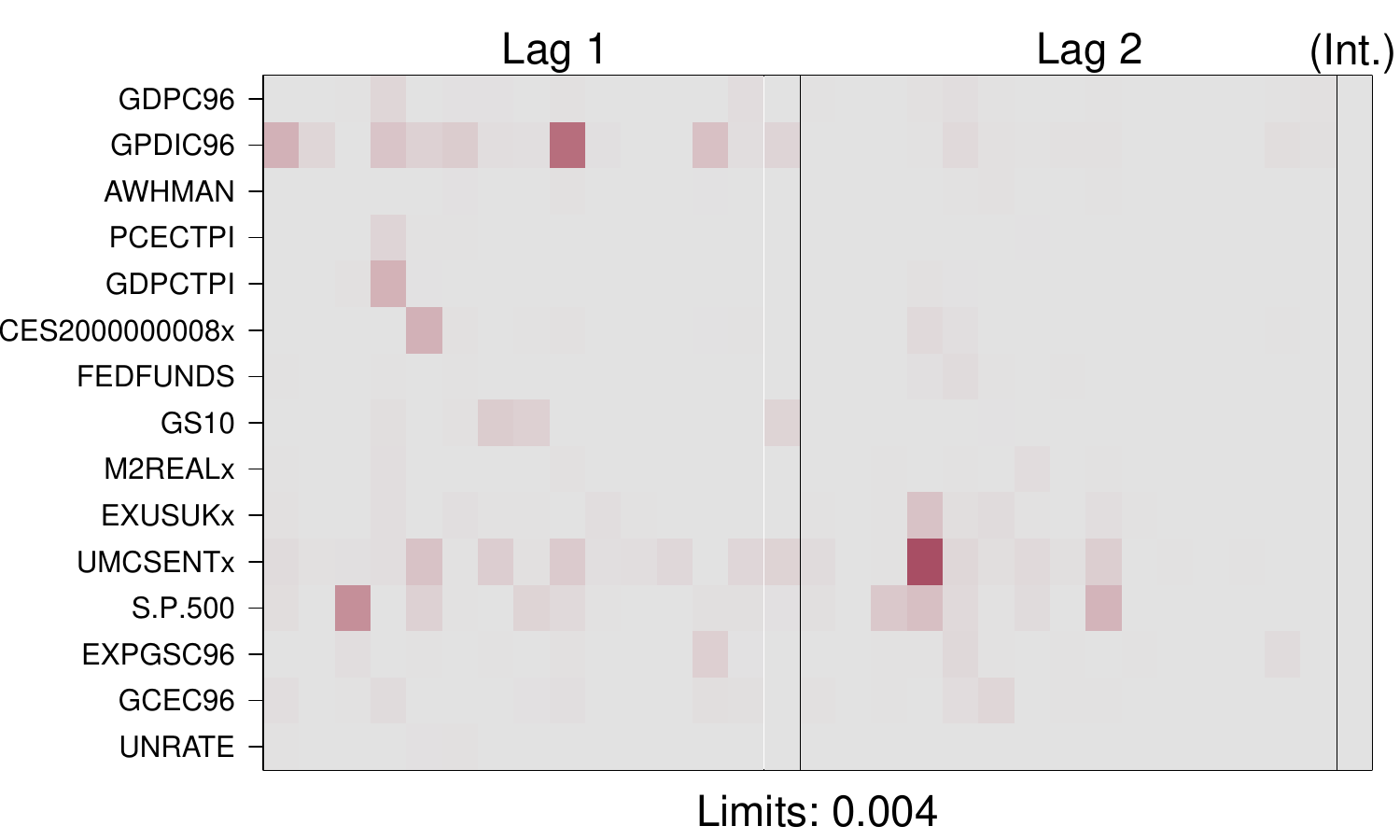}}\\
\caption{Posterior means in the large model -- USA.} \label{fig:heat_us}
\end{figure}

Before we turn to the forecasting exercise, we assess the amount of sparsity induced by our two proposed global-local shrinkage specifications, labeled TVP-SV NG and TVP-SV DL.
This analysis is based on inspecting heatmaps that show the posterior mean of the coefficients as well as the posterior mean of the standard deviations that determine the amount of time variation in the dynamic regression coefficients. 
Figs.\ \ref{fig:heat_ea} to \ref{fig:heat_us} show the corresponding heatmaps.  Red and blue squares indicate positive and negative values, respectively. 
To permit comparability, we use the same scaling across priors within a given country.

We start by  inspecting posterior sparsity attached to the time-invariant part of the models, provided in the upper panels of Figs.\ \ref{fig:heat_ea} to \ref{fig:heat_us}.  We generally find that the first own lag of a given variable appears to be important, while the second lag is slightly less important in most equations.  This can be seen by dense (i.e., colored) main diagonals elements. 
Turning to variables along the off-diagonal elements, i.e., the coefficients associated with variables $j \neq i$ in equation $i$, we find considerable evidence that the (un)employment rate as well as  long-term interest rates appears to load heavily on the other quantities in most country models,  as indicated by relatively dense columns associated with the first lag of unemployment and interest rates.

Equations that are characterized by a large amount of non-zero coefficients (i.e., dense rows) are mostly related to financial variables, namely, exchange rates, equity, and commodity prices. These observations are general in nature and relate to all three countries considered. 

In the next step,  we investigate sparsity in terms of the degree of time variation of the VAR coefficients (see the lower panels of Figs.\ \ref{fig:heat_ea} to \ref{fig:heat_us}). Here, we observe that, consistent with the dense pattern in $\boldsymbol{a}_0$, equations associated with financial variables display the largest amount of time variation. Interestingly, the results suggest that coefficients in the EA tend to display a greater propensity to drift as compared to the coefficients of the UK country model. 

Comparing the degree of shrinkage between both the DL and the NG prior reveals that the latter specification induces much more sparsity in large dimensional systems. While both priors yield rather sparse models, the findings point toward a much stronger degree of shrinkage of the NG prior. Notice that the NG prior also favors constant parameter specifications. This suggests that in large scale applications, the NG prior might be particular useful when issues of overparametrization are more of a concern, while in smaller models, the flexibility of the DL prior might be beneficial.

\subsection{Forecasting results} \label{forecast}

In this section, we examine the forecasting performance of the proposed prior specifications. The forecasting setup largely follows \cite{Huber2017} and focuses on the one-quarter- and one-year-ahead forecast horizons and three different information sets: small (3 variables), medium (7 variables), and large (15 variables).  We use an expanding window and a hold-out sample of 80 quarters, which results into the following hold out samples: 1995Q4-2015Q3 for the EA, 1997Q1-2016Q4 for the UK and 1995Q4-2015Q3 for the USA. 

\hl{Forecasts are evaluated using  log predictive scores (LPSs), a widely used metric to measure density forecast accuracy \citep[see, e.g.,][]{geweke2010comparing}. The LPSs are closely related to the marginal likelihood, a standard Bayesian measure for discriminating between competing models. It is also worth stressing that, as opposed to focusing on point forecasts, considering LPSs allows us to factor in how well a given model captures higher-order features of the predictive distribution. These higher-order features are important if the researcher is interested in reporting functions of the predictive distribution, such as growth at risk \citep{adrian2019vulnerable} or other measures relevant for capturing tail risks to macroeconomic outcomes \citep{clark2023tail}. Last, relying on point forecasts requires the researcher to choose a loss function (in addition to model and prior choice) in order to evaluate their quality, whereas using LPSs amounts to choosing the model-implied loss function.}

\begin{figure}[t]
\subfigure{%
\includegraphics[width=.329\textwidth, trim=25 45 30 45, clip, page=1]{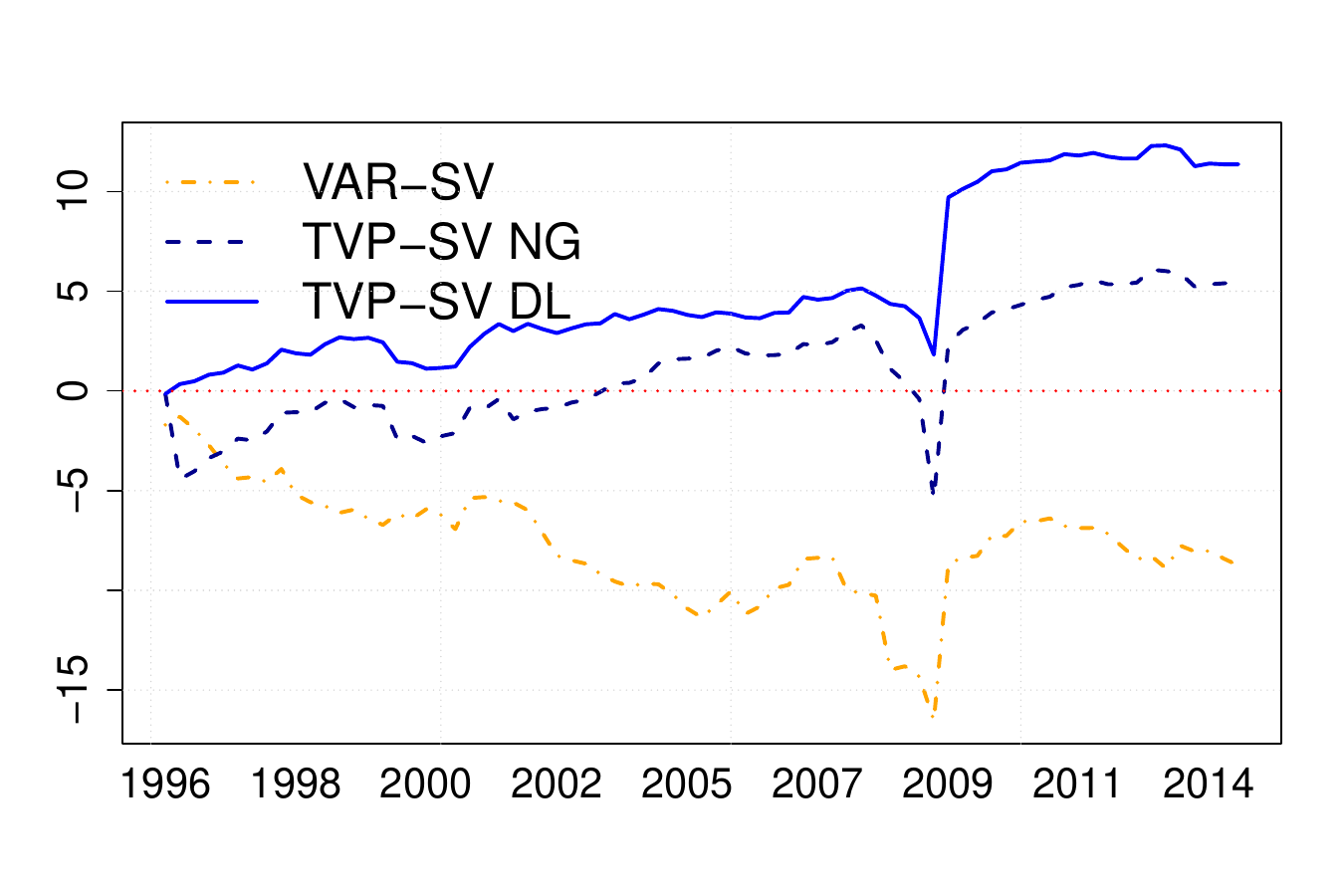}}
\subfigure{%
\includegraphics[width=.329\textwidth, trim=25 45 30 45, clip, page=4]{outputinflationIR.pdf}}
\subfigure{%
\includegraphics[width=.329\textwidth, trim=25 45 30 45, clip, page=7]{outputinflationIR.pdf}}\\
\subfigure{%
\includegraphics[width=.329\textwidth, trim=25 45 30 45, clip, page=2]{outputinflationIR.pdf}}
\subfigure{%
\includegraphics[width=.329\textwidth, trim=25 45 30 45, clip, page=5]{outputinflationIR.pdf}}
\subfigure{%
\includegraphics[width=.329\textwidth, trim=25 45 30 45, clip, page=8]{outputinflationIR.pdf}}\\
\subfigure{%
\includegraphics[width=.329\textwidth, trim=25 45 30 45, clip, page=3]{outputinflationIR.pdf}}
\subfigure{%
\includegraphics[width=.329\textwidth, trim=25 45 30 45, clip, page=6]{outputinflationIR.pdf}}
\subfigure{%
\includegraphics[width=.329\textwidth, trim=25 45 30 45, clip, page=9]{outputinflationIR.pdf}}\\
\caption{One-quarter-ahead cumulative log predictive Bayes factors \hl{for the three focus variables GDP growth, inflation, and short-term interest rates} over time, relative to the TVP-SV-VAR without shrinkage. Top row: Small model (3 variables). Middle row: Medium model (7 variables). Bottom row: Large model (15 variables). \hl{Left column: euro area (EA). Middle column: United Kingdom (UK). Right column: United States (US).}}\label{fig:lps1}
\end{figure}

We compare the NG and DL specifications with a simpler constant parameter Bayesian VAR (BVAR-SV) and a time-varying parameter VAR with a loose prior setting (TVP-SV) as a general benchmark. Specifically, this benchmark model assumes that the prior on $\sqrt{\omega}_j$ is given by
\begin{equation}
{\omega}_j \sim \mathcal{G}(1/2, 1/2) \Leftrightarrow \pm \sqrt{\omega}_j \sim \mathcal{N}(0, 1).
\end{equation}
On $\boldsymbol{a}_0$ and for the BVAR-SV, we use the NG shrinkage prior described in Section 3. 
 For the evaluation, we focus on the joint predictive distribution of three focal variables, namely, GDP growth, inflation, and short-term interest rates. This allows us to assess the predictive differences obtained by switching from small to large information sets.
\autoref{fig:lps1} summarizes the results for the one-step-ahead forecast horizon. All panels display LPSs for the three focus variables relative to the TVP-SV specification. To assess the overall forecast performance over the hold-out sample, particularly consider the rightmost point in the respective figures. 

Doing so reveals that the time-varying parameter specifications, TVP-SV NG and TVP-SV DL outperform the benchmark for all three countries and information sets as indicated by positive log predictive Bayes factors. Except for the EA and the small information set, this finding holds also true for the constant parameter VAR-SV specification. Zooming in and looking at performance differences among the priors reveals that the TVP-SV DL specification dominates in the case of small models.  The TVP-SV NG prior ranks second, and the constant parameter VAR-SV model performs worst. The dominance of the DL prior stems from the performance during the period of the global financial crisis 2008/2009. While predictions from all model specifications worsen,  they  deteriorate the least for the DL specification. In particular for the EA and the UK, the dominance of the DL prior stems mainly from improved forecast for short-term interest rates, see  Figs.~\ref{fig:marg_lps_ea1} and \ref{fig:marg_lps_uk1} in Appendix \ref{univLPS}.

It is worth noting that in small-dimensional models, the TVP-SV specification also performs quite well and proves to be a competitive alternative relative to the BVAR-SV model. This is due to the fact that parameters are allowed to move significantly with only little punishment introduced through the prior, effectively controlling for structural breaks and sharp movements in the underlying structural parameters.  This result corroborates findings in \cite{d2013macroeconomic} and appears to support  our conjecture that for small information sets, allowing for time variation proves to dominate the detrimental effect of the large number of additional parameters to be estimated.

In the next step,  we enlarge the information set and turn our focus to the seven variable VAR specifications. Here,  the picture changes slightly, and the NG prior outperforms forecasts of its competitors. Depending on the country, either forecasts of the DL specification or the constant parameter VAR-SV model rank second. For US data, it pays off to use a time-varying parameter specification since -- as with the small information set -- the BVAR-SV model performs worst.
Finally,  we turn to the large VAR specifications, featuring 15 variables. Here, we see a very similar picture as with the seven variable specification. The TVP-SV NG prior yields the best forecasts, with the constant parameter model turning out to be a strong competitor. Only for US data, both time-varying parameter specifications clearly outperform the constant parameter competitor. 

\begin{figure}[tp]
\subfigure{%
\includegraphics[width=.329\textwidth, trim=25 45 30 45, clip, page=1]{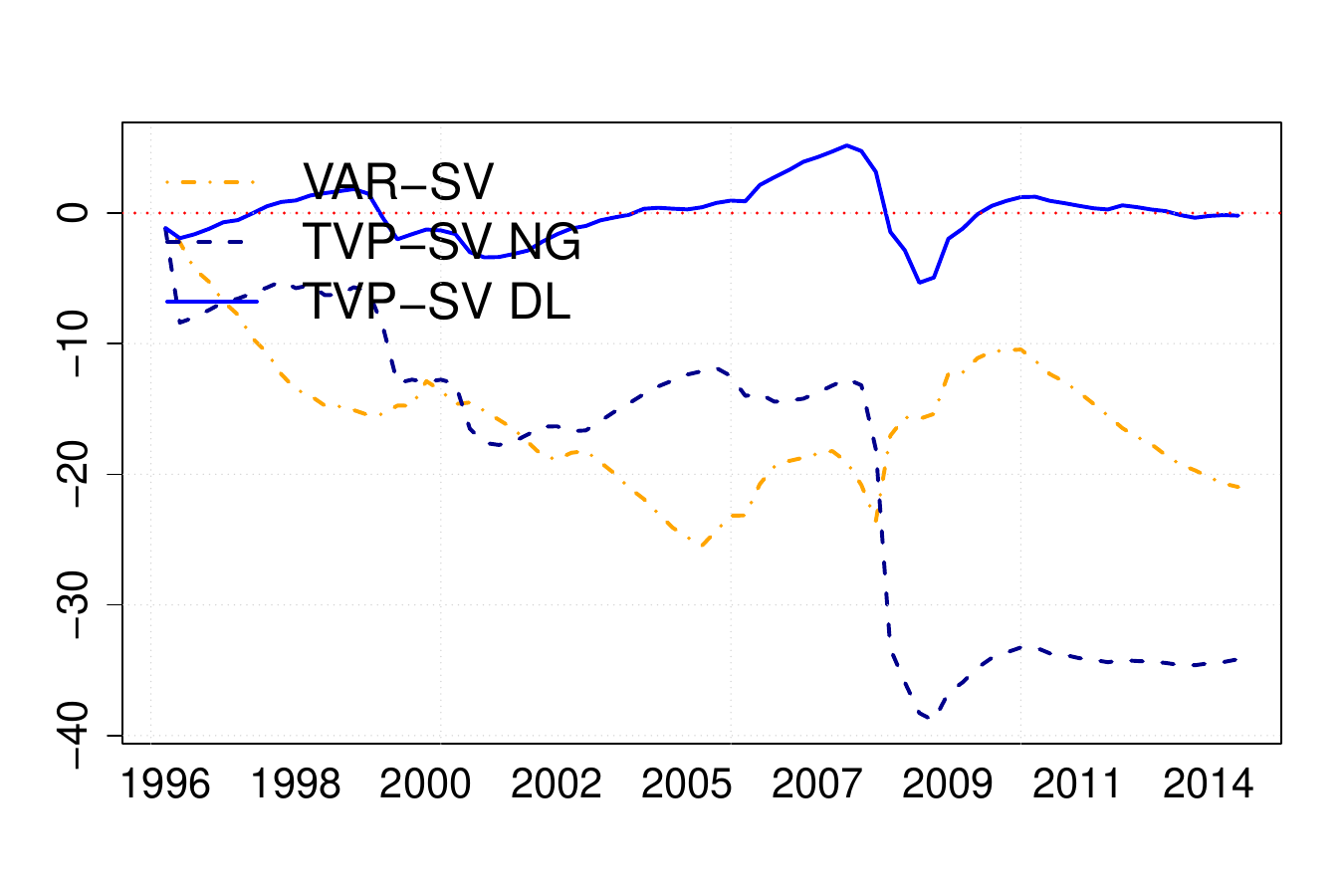}}
\subfigure{%
\includegraphics[width=.329\textwidth, trim=25 45 30 45, clip, page=4]{outputinflationIR4step.pdf}}
\subfigure{%
\includegraphics[width=.329\textwidth, trim=25 45 30 45, clip, page=7]{outputinflationIR4step.pdf}}\\
\subfigure{%
\includegraphics[width=.329\textwidth, trim=25 45 30 45, clip, page=2]{outputinflationIR4step.pdf}}
\subfigure{%
\includegraphics[width=.329\textwidth, trim=25 45 30 45, clip, page=5]{outputinflationIR4step.pdf}}
\subfigure{%
\includegraphics[width=.329\textwidth, trim=25 45 30 45, clip, page=8]{outputinflationIR4step.pdf}}\\
\subfigure{%
\includegraphics[width=.329\textwidth, trim=25 45 30 45, clip, page=3]{outputinflationIR4step.pdf}}
\subfigure{%
\includegraphics[width=.329\textwidth, trim=25 45 30 45, clip, page=6]{outputinflationIR4step.pdf}}
\subfigure{%
\includegraphics[width=.329\textwidth, trim=25 45 30 45, clip, page=9]{outputinflationIR4step.pdf}}
\caption{Four-quarter-ahead cumulative log predictive Bayes factors \hl{for the three focus variables GDP growth, inflation, and short-term interest rates} over time, relative to the TVP-SV-VAR with loose shrinkage. Top row: Small model (3 variables). Middle row: Medium model (7 variables). Bottom row: Large model (15 variables). \hl{Left column: euro area (EA). Middle column: United Kingdom (UK). Right column: United States (US).}}\label{fig:lps4}
\end{figure}

We now briefly examine forecasts for the four quarter horizon displayed in \autoref{fig:lps4}. For the small and medium-sized models, all competitors yield forecasts that are close or worse compared to the loose shrinkage benchmark prior model. The high degree of shrinkage induced by the NG prior yields particularly poor forecasts, especially for observations that fall in the period of the global financial crisis. The picture slightly reverses when considering the large-scale models. Here, all competitors easily outperform forecasts of the loose benchmark model, implying that shrinkage pays off. Viewed over all settings, the DL prior does a fine job in balancing the degree of shrinkage across model sizes. 

\subsection{Improving predictions through dynamic model selection} \label{pool}
The discussion in the previous subsection highlighted the marked heterogeneity of model performance over time. In terms of achieving superior forecasting results, one could ask whether there are gains from
dynamically selecting models.

Following \cite{raftery2010online, koop2012forecasting, onorante2016dynamic} we perform dynamic model selection by computing a set of weights for each model within a given model size. These weights are based on the predictive likelihood for the three focus variables at $t-1$. Intuitively speaking, this combination scheme implies that if a given model performed well in predicting last quarter's output, inflation and interest rates, it receives a higher weight in the next period. By contrast, models that performed badly receive less weight in the model pool. We further employ a so-called forgetting factor that induces persistence in the model weights over time. This implies that the weights are not only shaped by the most recent forecast performance of the underlying models, but also by their historical  forecasting performance. Finally, to select a given model, we simply pick the one with the highest weight.

The predicted weight associated with model $i$ is computed as follows 
\begin{equation}
 \mathfrak{w}_{t|t-1, i} := \frac{\mathfrak{w}^\alpha_{t-1|t-1, i}}{\sum_{i \in \mathcal{M}} \mathfrak{w}^\alpha_{t-1|t-1, {i}}},
\end{equation}
with $\alpha=0.99$ denoting a forgetting factor close to unity and $\mathfrak{w}_{t-1|t-1, i}$ is given by
\[
 \mathfrak{w}_{t-1|t-1, i} = \frac{\mathfrak{w}_{t-1|t-2,i}p_{t-1|t-2,i}}{\sum_{i \in \mathcal{M}} \mathfrak{w}_{t-1|t-2,i}p_{t-1|t-2,i}}.
\]
Here, $p_{t-1|t-2,i}$ denotes the one-step-ahead predictive likelihood for the three focus variables in $t-1$ for model $i$ within the model space $\mathcal{M}$. Letting $t_0$ stand for the final quarter of the training sample, the initial weights $\mathfrak{w}_{t_0+1|t_0,i}$ are assumed to be equal for each model.

\begin{figure}[tp]
\subfigure{%
\includegraphics[width=.329\textwidth, trim=25 45 30 45, clip, page=1]{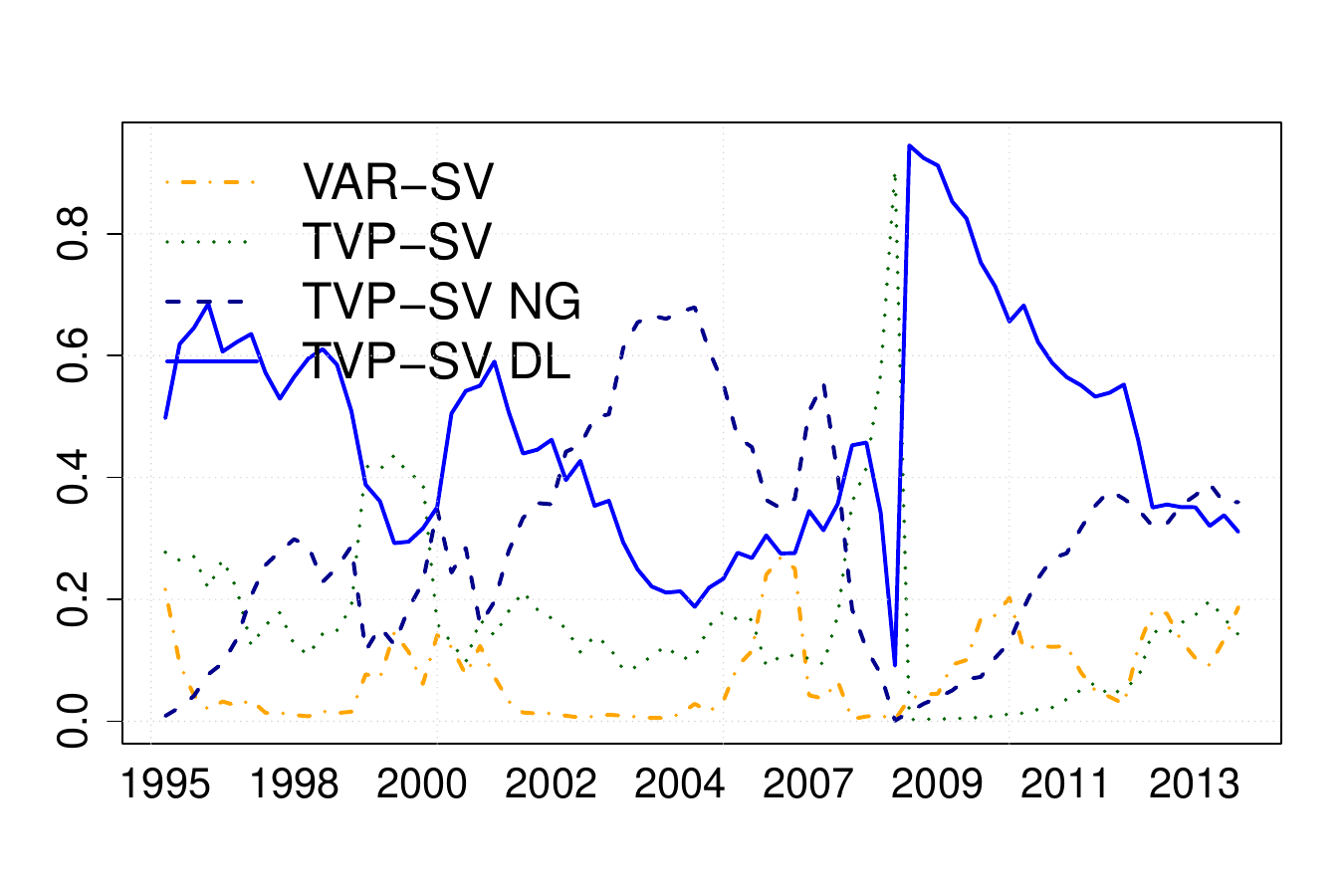}}
\subfigure{%
\includegraphics[width=.329\textwidth, trim=25 45 30 45, clip, page=4]{Jointweights.pdf}}
\subfigure{%
\includegraphics[width=.329\textwidth, trim=25 45 30 45, clip, page=7]{Jointweights.pdf}}\\
\subfigure{%
\includegraphics[width=.329\textwidth, trim=25 45 30 45, clip, page=2]{Jointweights.pdf}}
\subfigure{%
\includegraphics[width=.329\textwidth, trim=25 45 30 45, clip, page=5]{Jointweights.pdf}}
\subfigure{%
\includegraphics[width=.329\textwidth, trim=25 45 30 45, clip, page=8]{Jointweights.pdf}}\\
\subfigure{%
\includegraphics[width=.329\textwidth, trim=25 45 30 45, clip, page=3]{Jointweights.pdf}}
\subfigure{%
\includegraphics[width=.329\textwidth, trim=25 45 30 45, clip, page=6]{Jointweights.pdf}}
\subfigure{%
\includegraphics[width=.329\textwidth, trim=25 45 30 45, clip, page=9]{Jointweights.pdf}}
\caption{Model weights over time as implied by the predictive performance \hl{for the three focus variables GDP growth, inflation, and short-term interest rates}. Top row: Small model (3 variables). Middle row: Medium model (7 variables). Bottom row: Large model (15 variables). \hl{Left column: euro area (EA). Middle column: United Kingdom (UK). Right column: United States (US).}}
\label{fig:dynMod2}
\end{figure}

Before proceeding to the forecasting results, \autoref{fig:dynMod2} shows the model weights over time.  One interesting regularity for small-scale models is that especially during the crisis period, the algorithm selects the benchmark, weak shrinkage TVP-SV model. This choice, however, proves to be of transient nature and the algorithm quickly adapts and switches back to either the TVP-SV NG or the TVP-SV DL model. We interpret this finding to be related to the necessity to quickly adjust to changes in the underlying macroeconomic conditions in light of the small information set adopted. The TVP-SV model allows for large shifts in the underlying regression coefficients, whereas the specifications based on hierarchical shrinkage priors introduce shrinkage, which excels over the full hold-out period but proves to be detrimental during crisis episodes.

For the medium and large information set, the model weights corroborate the results reported in the previous subsection. Specifically, we see that TVP-SV NG  and TVP-SV DL receive high weights during the global financial crisis, while the BVAR-SV receives large shares of posterior probability during the remaining periods. This implies that during periods with overall heightened uncertainty, gains from using a time-varying parameter framework are sizable. 

\begin{figure}[tp]
\subfigure{%
\includegraphics[width=.329\textwidth, trim=25 45 30 50, clip, page=1]{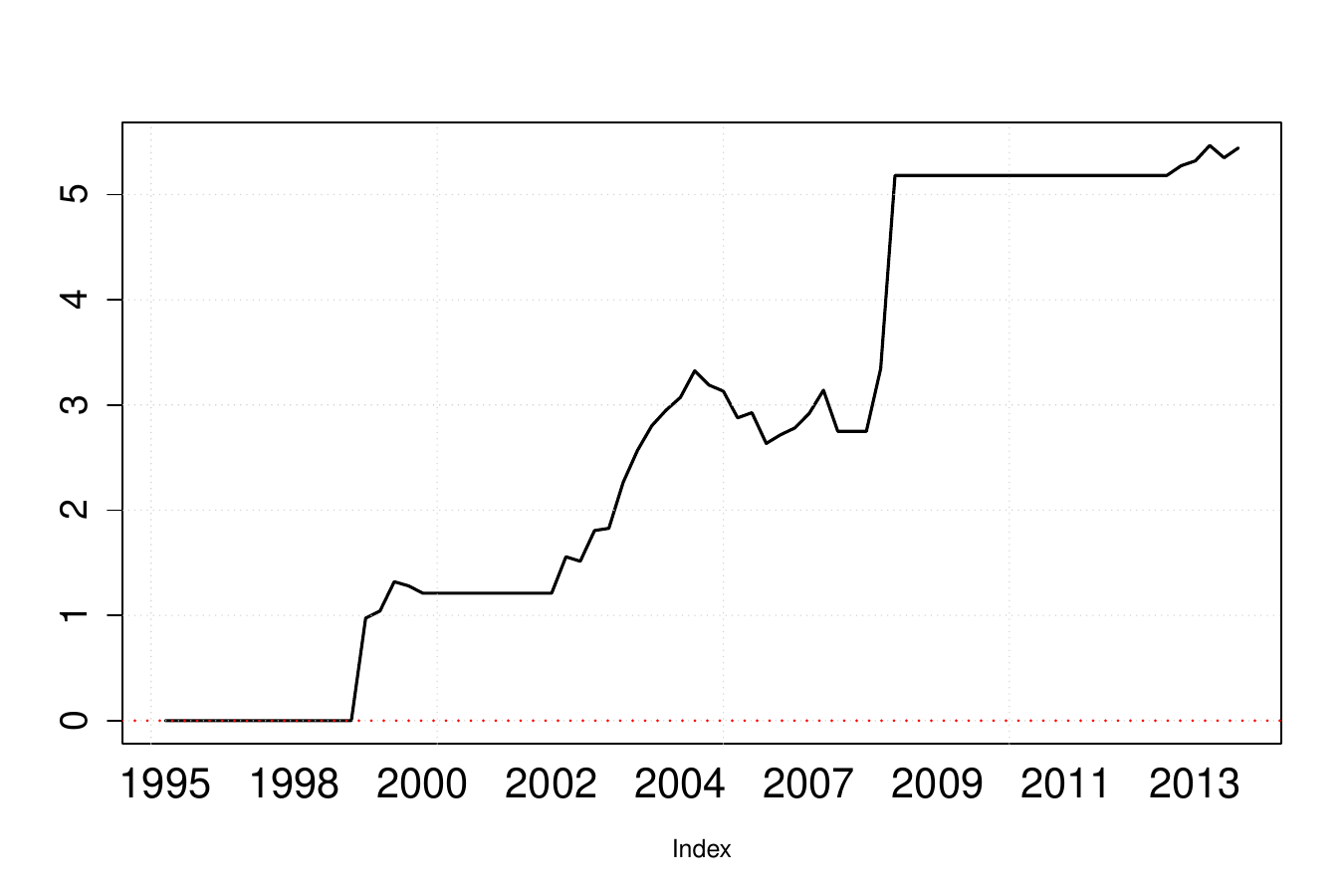}}
\subfigure{%
\includegraphics[width=.329\textwidth, trim=25 45 30 50, clip, page=4]{Jointcomb.pdf}}
\subfigure{%
\includegraphics[width=.329\textwidth, trim=25 45 30 50, clip, page=7]{Jointcomb.pdf}}\\
\subfigure{%
\includegraphics[width=.329\textwidth, trim=25 45 30 50, clip, page=2]{Jointcomb.pdf}}
\subfigure{%
\includegraphics[width=.329\textwidth, trim=25 45 30 50, clip, page=5]{Jointcomb.pdf}}
\subfigure{%
\includegraphics[width=.329\textwidth, trim=25 45 30 50, clip, page=8]{Jointcomb.pdf}}\\
\subfigure{%
\includegraphics[width=.329\textwidth, trim=25 45 30 50, clip, page=3]{Jointcomb.pdf}}
\subfigure{%
\includegraphics[width=.329\textwidth, trim=25 45 30 50, clip, page=6]{Jointcomb.pdf}}
\subfigure{%
\includegraphics[width=.329\textwidth, trim=25 45 30 50, clip, page=9]{Jointcomb.pdf}}
\caption{Log predictive Bayes factors of dynamic model selection relative to the best performing model over time, \hl{as implied by the predictive performance for the three focus variables GDP growth, inflation, and short-term interest rates.} Top row: Small model (3 variables). Middle row: Medium model (7 variables). Bottom row: Large model (15 variables). \hl{Left column: euro area (EA). Middle column: United Kingdom (UK). Right column: United States (US).}}
\label{fig:dynMod1}
\end{figure}

We now turn to the forecasting results using DMS, provided in \autoref{fig:dynMod1}. The figure shows the log predictive Bayes factors relative to the best performing models over the whole sample period. These correspond to those achieving the highest cumulative log predictive Bayes factors in  \autoref{fig:lps1}. 

The results indicate that dynamic model selection tends to improve forecasts throughout all model sizes and for all three country data sets. In particular, during the period of the global financial crisis, selecting from a pool of model pays off. Forecast gains during the crisis are more pronounced for the EA and the UK, whereas with US data, forecasts are more gradually improving over the sample period. Forecasts for the EA that are based on the large information set are less precise during the period from 2000 to 2012 compared to the benchmark models. This might be related to the creation of the euro, which in turn has triggered a fundamental shift in the joint dynamics of the EA's macro model. Due to the persistence in the models' weights, the model selection algorithm takes some time to adapt to the new regime. This can be seen by investigating the latest period in the sample, in which dynamic model selection again outperforms forecasts of the benchmark model. In other words, for EA data, either restricting the sample period to post 2000 or reducing the persistence via the forgetting factors might improve forecasting results. 

\section{Conclusive remarks}
In this paper, our goal is to investigate how model complexity (loosely defined as the dimension of the state space) interacts with model size. The initial conjecture is that small but flexible (TVP) models are capable of achieving predictive performances that are comparable to large-scale models that assume constant parameters (but time-varying error variances). In three different data sets, we find precisely this pattern. The larger the model gets, the less we require time-varying parameters and the closer the forecasting performance of the large TVP-VAR is to the constant parameter large VAR with SV.

This is mainly driven by the fact that modern global-local shrinkage priors introduce substantial shrinkage on the process innovation variances and the degree of shrinkage increases rapidly in the number of coefficients. In case that the model is small, the TVPs control for omitted variable biases, and this proves to be a good substitute for using larger data sets in combination with simpler models. 

\section*{Data availability statement}
The data that support the findings of this study are openly available. A detailed description is given in Appendix \ref{app:data}.

\begin{funding}
Luis Gruber, Florian Huber, and Gregor Kastner acknowledge funding from the Austrian Science Fund (FWF) for the project ``High-dimensional statistical learning: New methods to advance economic and sustainability policy'' (ZK 35), jointly carried out by the University of Klagenfurt, Paris Lodron University Salzburg, TU Wien, and the Austrian Institute of Economic Research (WIFO).
\end{funding}


\bibliographystyle{imsart-nameyear}
\bibliography{bibtex/favar.bib,bibtex/mpShocks.bib}

\clearpage

\begin{appendix}
\section{Data overview} \label{app:data}
\begin{table}[h]
\caption{Data overview}\label{tbl:data}
\resizebox{\columnwidth}{!}{%
\begin{tabular}{llllcccc}
\textbf{Variable}	&	\textbf{EA}	&	\textbf{UK}	&	\textbf{US}	&	\textbf{m=3}	&	\textbf{m=7}	&	\textbf{m=15}	&	\textbf{T-code}	\\
\midrule															
\textit{Real gross domestic product}	&	YER,  GDP at market prices, 	&	GDP at market prices.	&	GDPC96, bn of chained 2009 USD.	&	x	&	x	&	x	&	1	\\
	&	chain linked volumes, sa.	&		&		&		&		&		&		\\
\textit{Prices}	&	YED, GDP deflator index.	&	GDP deflator at market prices.	&	GDPCTPI, chain-type price index.	&	x	&	x	&	x	&	1	\\
\textit{Short-term interest rates}	&	STN, 3-month euribor in \% p.a.	&	Bank rate, avg. of monthly series.	&	FEDFUNDS, eff. federal funds rate in \%.	&	x	&	x	&	x	&	2	\\
\midrule															\\
\textit{Investment}	&	ITR, Gross fixed capital formation, 	&	Gross fixed capital formation,	&	GPDIC96, bn of chained 2009 USD.	&		&	x	&	x	&	1	\\
	&	chain linked volume, sa.	&	 chained volume measure.	&		&		&		&		&		\\
\textit{Consumption}	&	PCR, individual consumption expenditure, 	&	Household consumption, 	&	PCECTPI, personal consumption 	&		&	x	&	x	&	1	\\
	&	chain linked volume, sa.	&	chained volume measure.	&	expenditures: chain-type price index.	&		&		&		&		\\
\textit{Exchange rate}	&	EEN, nominal eff. exchange rate 	&	Nominal dollar per pound 	&	EXUSUKx, dollar pound 	&		&	x	&	x	&	1	\\
	&	vis-a-vis 38 non-euro area trading partners.	&	exchange rate.	&	foreign exchange rate.	&		&		&		&		\\
\textit{Employment / Unemployment}	&	URX, percentage of civilian workforce, sa.	&	Employment in heads.	&	UNRATE, civilian unempl. rate in \%.	&		&	x	&	x	&	2	\\
\midrule															
\textit{Wages}	& 	WRN, wages per head.	&	Spliced average weekly 	&	CES2000000008x, real avg.  	&		&		&	x	&	1	\\
	& 		& 	earnings series.	& 	hourly earnings.	& 		& 		& 	x	& 		\\
\textit{Money}	& 	M2, sa, OECD data.	&	Stock of ``broad money'',	&	M2REALx, real M2 money stock	&		&		&	x	&	1	\\
	&		&	break-adjusted.	&	 (bn of 1982-84 USD), deflated by CPI.	&		&		&		&		\\
\textit{Equity prices}	&	Equity share price index,	&	Spliced monthly share price index 	&	S\&P 500 Common composite	&		&		&	x	&	1	\\
	&	OECD data.	&	weighted by market capitalization.	&	 stock price index.	&		&		&		&		\\
\textit{Long-term interest rate}	&	LTN, 10-year gov. bond yields in \% p.a.	&	10-year gov. bond yields in \% p.a.	&	GS10, 10-year gov. Bond yields in \%.	&		&		&	x	&	2	\\
\textit{Government consumption}	&	GCR, general gov. expenditure, 	&	Gov. consumption, 	&	GCEC96, real gov. spending,	&		&		&	x	&	1	\\
	&	chain linked volumes, sa.	&	chained volume measure.	&	in bn of 2009 chained USD.	&		&		&		&		\\
\textit{Exports}	&	XTR, exports of goods and services, 	&	Export volumes, 	&	EXPGSC96, bn of chained 2009 USD.	&		&		&	x	&	1	\\
	&	chain linked volumes, sa.	&	chained volume measures.	&		&		&		&		&		\\
\midrule															
\textit{Commodity price index}	&	COMPR, weighted sum of oil prices and 	&		&		&		&		&	x	&	1	\\
	&	non-oil commodity prices in US dollars.	&		&		&		&		&		&		\\
\textit{Labour productivity}	&	LPROD, ratio of real GDP and total employment.	&		&		&		&		&	x	&	1	\\
\textit{Mortgage rate}	&		&	Household var. mortg. rate in \% p.a.	&		&		&		&	x	&	2	\\
\textit{Corporate bond spread}	&		&	Spliced interpolated  corporate	&		&		&		&	x	&	2	\\
	&		&	bond spreads.	&		&		&		&		&		\\
\textit{Hours worked}	&		&		&	AWHMAN, avg weekly hours of production 	&		&		&	x	&	1	\\
	&		&		&	and nonsup. employees: manufacturing.	&		&		&		&		\\
\textit{Consumer sentiment}	&		&		&	UMCSENTx, consumer sentiment index.	&		&		&	x	&	1	\\
\bottomrule															
	\multicolumn{8}{p{1.85\columnwidth}}{\footnotesize{\textbf{Notes:} Data are obtained from \url {https://eabcn.org/page/area-wide-model} for the euro area, from \url{https://www.bankofengland.co.uk/statistics/research-datasets} for the UK and from \url{https://research.stlouisfed.org/econ/mccracken/fred-databases/} for the USA. For the euro area, equity prices and M2 obtained from \url{https://data.oecd.org/}; employment data for UK transformed using log differences. Transformation codes (T-codes) are as follows:
	1 -- log differences, 
	2 -- raw data.}}
\end{tabular}%
}
\label{tab:dataset}%
\end{table}

\clearpage

\section{Additional empirical results}
\label{univLPS}

\begin{figure}[h]
\subfigure[Real GDP: small model]{\includegraphics[width=.329\textwidth, trim=25 45 30 45, clip, page=1]{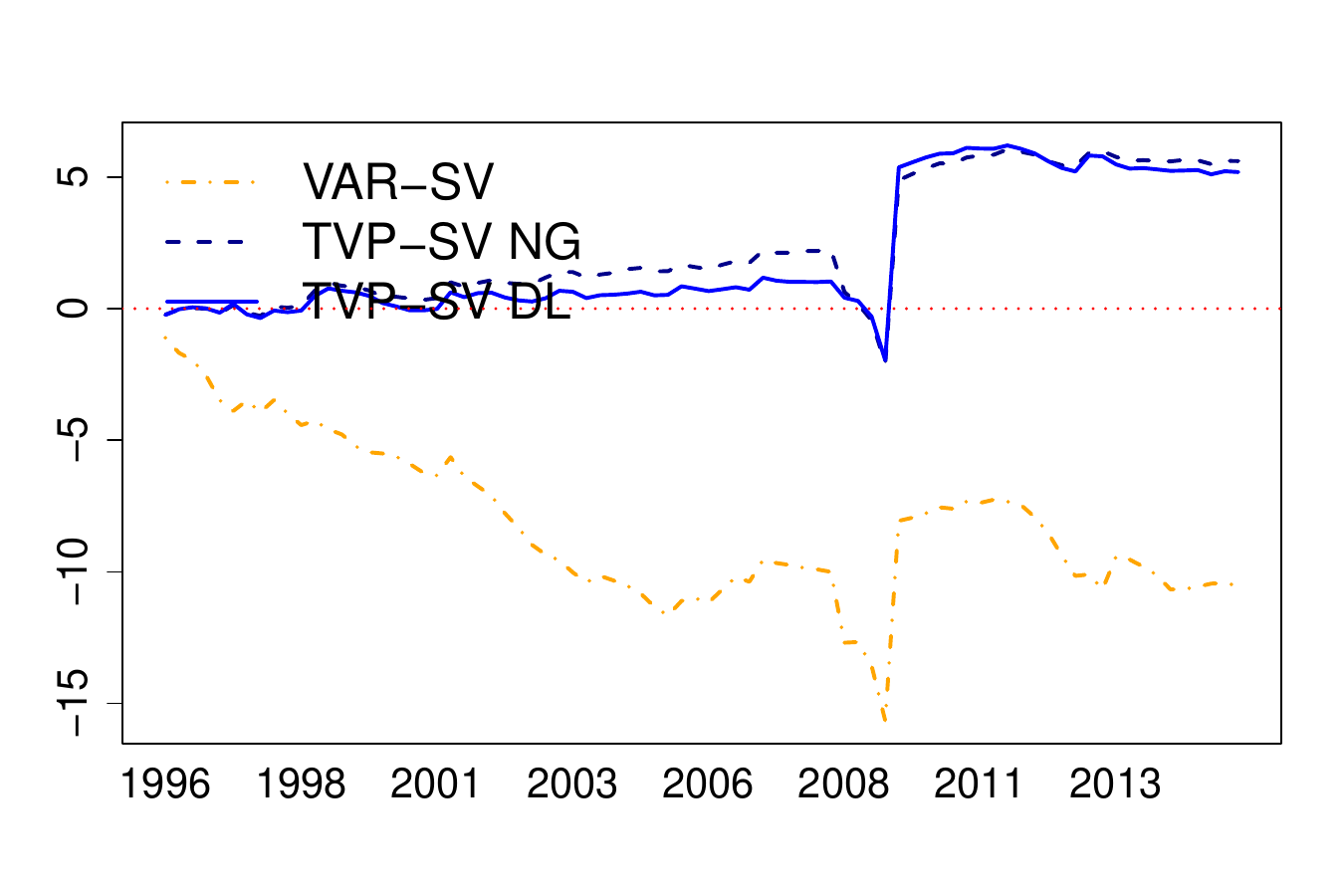}}
\subfigure[GDP deflator: small model]{\includegraphics[width=.329\textwidth, trim=25 45 30 45, clip, page=2]{EA_univ.pdf}}
\subfigure[Short rates: small model]{\includegraphics[width=.329\textwidth, trim=25 45 30 45, clip, page=3]{EA_univ.pdf}}\\
\subfigure[Real GDP: medium model]{\includegraphics[width=.329\textwidth, trim=25 45 30 45, clip, page=4]{EA_univ.pdf}}
\subfigure[GDP deflator: medium model]{\includegraphics[width=.329\textwidth, trim=25 45 30 45, clip, page=5]{EA_univ.pdf}}
\subfigure[Short rates: medium model]{\includegraphics[width=.329\textwidth, trim=25 45 30 45, clip, page=6]{EA_univ.pdf}}\\
\subfigure[Real GDP: large model]{\includegraphics[width=.329\textwidth, trim=25 45 30 45, clip, page=7]{EA_univ.pdf}}
\subfigure[GDP deflator: large model]{\includegraphics[width=.329\textwidth, trim=25 45 30 45, clip, page=8]{EA_univ.pdf}}
\subfigure[Short rates: large model]{\includegraphics[width=.329\textwidth, trim=25 45 30 45, clip, page=9]{EA_univ.pdf}}
\caption{euro area (EA): Univariate cumulative log predictive one-quarter-ahead Bayes factors over time relative to the TVP-SV-VAR with loose shrinkage.
}
\label{fig:marg_lps_ea1}
\end{figure}

\begin{figure}[p]
\subfigure[Real GDP: small model]{\includegraphics[width=.329\textwidth, trim=25 45 30 45, clip, page=1]{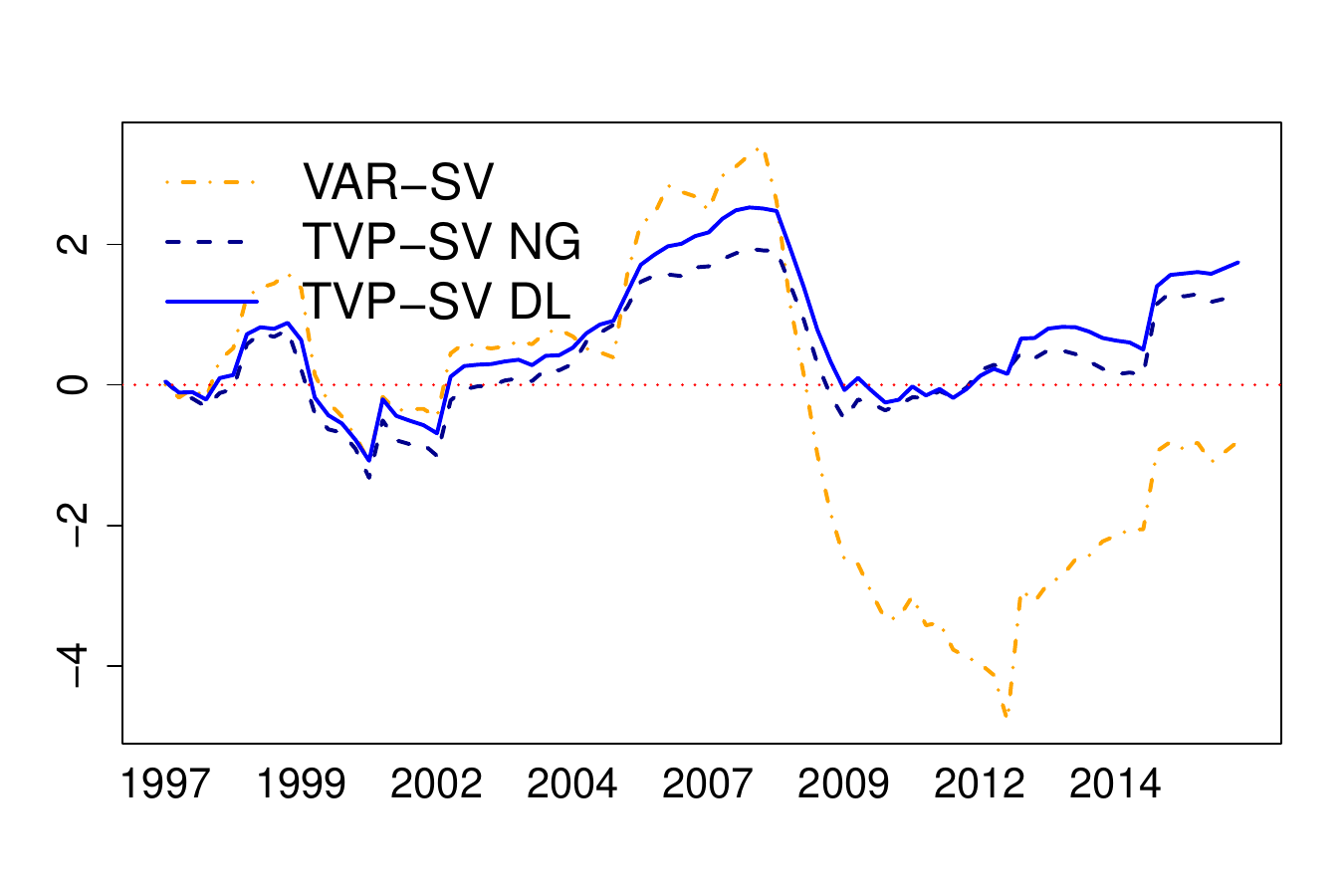}}
\subfigure[GDP deflator: small model]{\includegraphics[width=.329\textwidth, trim=25 45 30 45, clip, page=2]{UK_univ.pdf}}
\subfigure[Short rates: small model]{\includegraphics[width=.329\textwidth, trim=25 45 30 45, clip, page=3]{UK_univ.pdf}}\\
\subfigure[Real GDP: medium model]{\includegraphics[width=.329\textwidth, trim=25 45 30 45, clip, page=4]{UK_univ.pdf}}
\subfigure[GDP deflator: medium model]{\includegraphics[width=.329\textwidth, trim=25 45 30 45, clip, page=5]{UK_univ.pdf}}
\subfigure[Short rates: medium model]{\includegraphics[width=.329\textwidth, trim=25 45 30 45, clip, page=6]{UK_univ.pdf}}\\
\subfigure[Real GDP: large model]{\includegraphics[width=.329\textwidth, trim=25 45 30 45, clip, page=7]{UK_univ.pdf}}
\subfigure[GDP deflator: large model]{\includegraphics[width=.329\textwidth, trim=25 45 30 45, clip, page=8]{UK_univ.pdf}}
\subfigure[Short rates: large model]{\includegraphics[width=.329\textwidth, trim=25 45 30 45, clip, page=9]{UK_univ.pdf}}
\caption{United Kingdom (UK): Univariate cumulative log predictive one-quarter-ahead Bayes factors over time relative to the TVP-SV-VAR with loose shrinkage.
}
\label{fig:marg_lps_uk1}
\end{figure}

\begin{figure}[p]
\subfigure[Real GDP: small model]{\includegraphics[width=.329\textwidth, trim=25 45 30 45, clip, page=1]{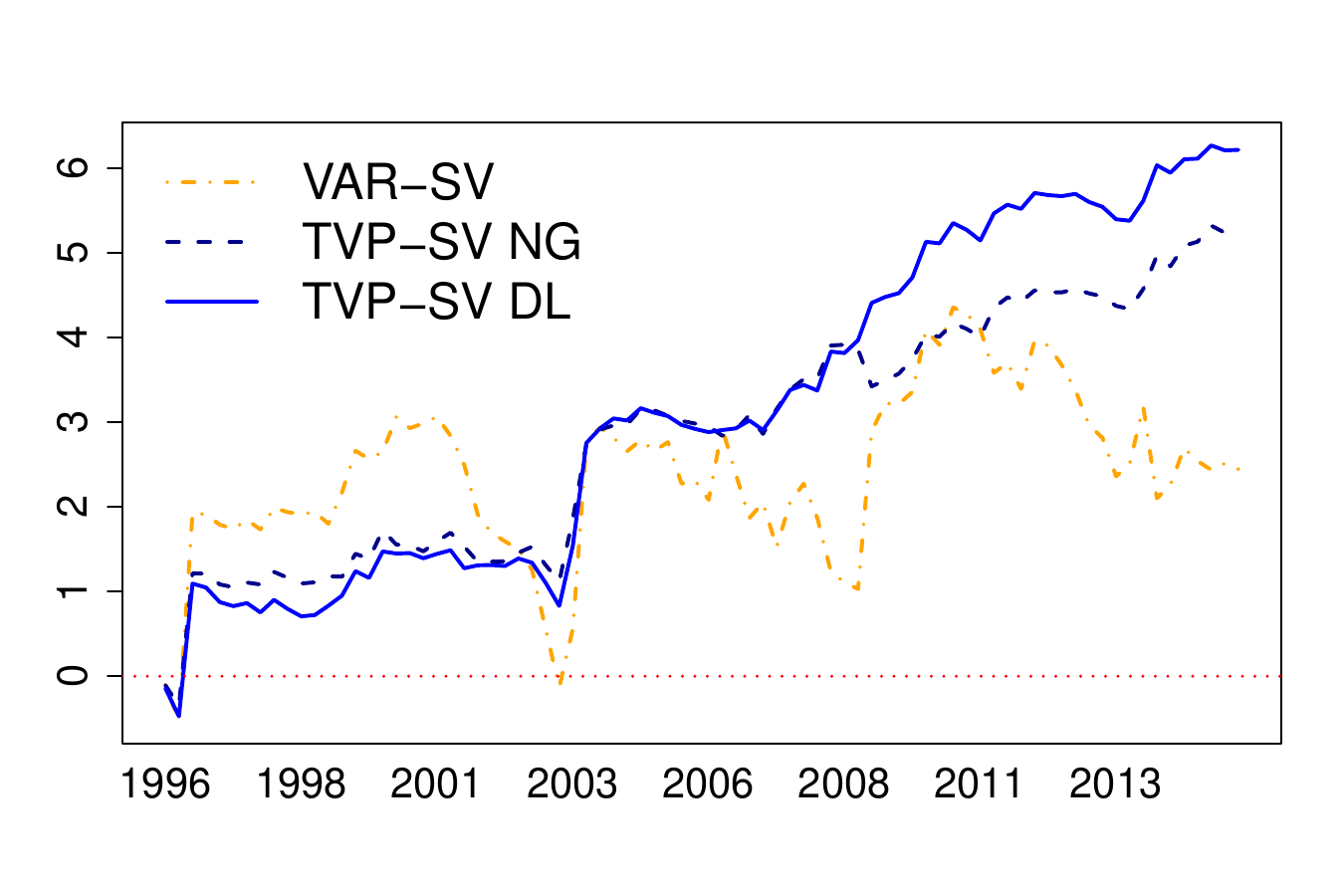}}
\subfigure[GDP deflator: small model]{\includegraphics[width=.329\textwidth, trim=25 45 30 45, clip, page=2]{US_univ.pdf}}
\subfigure[Short rates: small model]{\includegraphics[width=.329\textwidth, trim=25 45 30 45, clip, page=3]{US_univ.pdf}}\\
\subfigure[Real GDP: medium model]{\includegraphics[width=.329\textwidth, trim=25 45 30 45, clip, page=4]{US_univ.pdf}}
\subfigure[GDP deflator: medium model]{\includegraphics[width=.329\textwidth, trim=25 45 30 45, clip, page=5]{US_univ.pdf}}
\subfigure[Short rates: medium model]{\includegraphics[width=.329\textwidth, trim=25 45 30 45, clip, page=6]{US_univ.pdf}}\\
\subfigure[Real GDP: large model]{\includegraphics[width=.329\textwidth, trim=25 45 30 45, clip, page=7]{US_univ.pdf}}
\subfigure[GDP deflator: large model]{\includegraphics[width=.329\textwidth, trim=25 45 30 45, clip, page=8]{US_univ.pdf}}
\subfigure[Short rates: large model]{\includegraphics[width=.329\textwidth, trim=25 45 30 45, clip, page=9]{US_univ.pdf}}
\caption{United States (US): Univariate cumulative log predictive one-quarter-ahead Bayes factors over time relative to the TVP-SV-VAR with loose shrinkage.
}
\label{fig:marg_lps_us1}
\end{figure}

We display forecasting performances for three economies (EA, UK, US) and different model specifications for three key variables (real GDP, GDP deflator, short rates) in Figs.~\ref{fig:marg_lps_ea1} to \ref{fig:marg_lps_us1}.

\end{appendix}

\clearpage



\end{document}